# ФОТОЭМИССИЯ ИЗ МЕТАЛЛИЧЕСКИХ НАНОЧАСТИЦ

И.Е.Проценко, А.В.Усков

*Физический институт им. П.Н.Лебедева РАН, 119991, Москва, Ленинский Проспект, 53*

*ООО «Плазмоника» 109382, Москва, ул. Нижние поля 52/1.*

protsenk@sci.lebedev.ru; alexusk@sci.lebedev.ru

## Аннотация

Подход А.М.Бродского и Ю.Я.Гуревича обсуждатся и обобщается для фотоэмиссии из металлических наночастиц с учетом возбуждения в них локализованного плазмонного резонанса (ЛПР) и изменения электромагнитного поля (ЭМП) и массы электрона проводимости на границе металл-внешняя среда. Новым результатом является увеличение фотоэмиссионного тока из слоя наночастиц в несколько раз по сравнению со случаем сплошного слоя металла из-за увеличения интенсивности ЭМП вблизи поверхности частиц при возбуждении ЛПР а также относительно большой площади поверхности частиц и упомянутых граничных эффектов. Результаты могут быть использованы для создания фотодетекторов, фотопреобразователей, для дальнейшего изучения фотоэмиссии из наночастиц.

# 1. Введение.

Известно, что колебания электронной плотности металлических наночастиц имеют резонансную частоту в видимой части спектра или ближнем ИК диапазоне спектра – локализованный плазмонный резонанс (ЛПР). Исследованию оптических и электрофизических эффектов, связаных с возбуждением ЛПР и других подобных им резонансов в металлических наночастицах и наноструктурах посвящена быстро развивающаяся современная область фи-ики – наноплазмоника [4]. Наличие ЛПР вызвано поверхностным зарядом частиц [1-4]: границы частицы образуют "потенциальную яму" для колебаний электронной плотности частицы. Частота ЛПР для малых частиц (с характерным размером менее 40 нм) слабо зависит от размера частиц и сильнее – от их формы, металла и материала окружения. ЛПР может возбуждаться внешним электромагнитным полем (ЭМП). В случае ЛПР энергия запасается как в колебаниях электронной плотности, так и в самом ЭМП, связанном с этими колебаниями. Плотность энергии ЭМП на частоте ЛПР внутри и на расстоянии менее или порядка длины волны ЛПР около наночастицы оказывается примерно в Q раз выше [5], чем плотность энергии внешнего ЭМП, возбуждающего частицу, где Q - добротность ЛПР, зависящая от потерь, в т.ч. на рассеяние поля частицей во внешнее пространство и поглощение в материале частицы. В экспериментах обычно Q<10 [1-4;6,7], хотя теоретические оценки предсказывают максимальное значение Q в несколько десятков [8]. Наночастица при возбуждении ЛПР ведет себя как резонатор для ЭМП но, в отличии от, например, резонатора Фабри-Перо, в наночастице–«резонаторе» присутствует и поле ближней зоны, т.е. кулоновское поле, связанное с зарядами; ближняя зона находится на расстояниях $\leq \lambda$ от частицы, где $\lambda$ - длина волны ЭМП. Резонансные свойства металлических наночастиц (которые часто называют "плазмонными" [1-4;8]) и соответствующая способность частиц "концентрировать" вокруг и внутри себя электромагнитное поле позволили предсказать и наблюдать экспериментально ряд новых эффектов [1-4;8;9], наиболее известный из которых - гигантское комбинационное рассеяние [9]. Были предложены и, в некоторых случаях, реализованы новые оптоэлектронные устройства с плазмонными наночастицами, в том числе сенсоры [1-4;9;10,11], наноразмерные лазеры [12-15], солнечные фотопреобразователи повышенной эффективности [16-18].

Наночастицы, используемые в оптических устройствах, иногда рассматриваются как "наноантенны" [1-4;19,20]. При разработке и моделировании приборов, использующих оптические ("плазмонные") свойства металлических наночастиц, интересно и важно исследовать также электрофизические свойства контакта плазмонных наночастиц с их окружением, например контакта между частицей и поверхностью, на которой она находится. Перенос, носителей заряда (электронов или дырок) и (или) энергии из частицы в ее окружение и обратно может оказать существенное влияние на свойства устройства, включающего наночастицы, как в положительную, так и в отрицательную сторону, вплоть до полного уничтожения ЛПР и связанных с ним эффектов. Например, известен эффект усиления фотолюминесценции за счет добавления к люминесцирующим объектам (например, молекулам красителя в растворе) металлических наночастиц. Но в том случае, если расстояние между люминесцирующей молекулой и наночастицей достаточно мало (менее нескольких нанометров), происходит эффективная безызлучательная релаксация и люминесценция полностью подавляется [1-4,21]. Предложены схемы фотопреобразователей, в которых одновременно используются как оптические плазмонные, так и электрофизические свойства металлических наночастиц, например, для электрического соединения (коммутации) каскадов фотопреобразователя [22]. Есть данные о том, что эффективность солнечных фотопреобразователей повышается, как сказано в [23], при инжекции в полупроводниковую подложку носителей от электрон-дырочных пар фотоиндуцированных в металлических наночастицах, находящихся на подложке, при возбуждении в частицах ЛПР. Вместе с тем, процессы переноса носителей через контакты плазмонных наночастиц с подложкой, при возбуждении в них ЛПР, исследованы значительно меньше, чем оптические свойства наночастиц. Это связано со сложностью как теоретических, так и экспериментальных исследований электрофизических свойств наноразмерных объектов. Изучение транспортных и других электрофизических свойств контактов подложки с металлическими наночастицами, выполняемое одновременно с исследованием оптических свойств частиц, необходимо для создания эффективных плазмонных оптоэлектронных устройств, в том числе солнечных фотопреобразователей, фотодетекторов, наноразмерных светодиодов и нанолазеров.

Одним из эффектов, связанных с переносом носителей при возбуждении ЛПР наночастиц, является фотоэмиссия из наночастиц, теоретически исследуемая в настоящей работе. Фотоэмиссия из наночастиц может существенно отличаться от фотоэмиссии из больших (по сравнению с длиной волны ЭМП), "макроскопических" металлических структур (например, из слоев металла), так как, во-первых, ЭМП внутри частицы и вне ее, вблизи поверхности наночастицы, сильно возрастает при возбуждении ЛПР, во-вторых, отношение площади поверхности наночастицы к ее объему – существенно больше, чем у макроскопических структур. Последнее важно, так как основной вклад в фотоэмиссию вносит "поверхностный" фотоэффект, когда электрон поглощает фотон и покидает металл при столкновении с поверхностью металла [24]. Другой тип фотоэффекта - "объемный", когда фотон поглощается при столкновении электрона под поверхностью металла, вносит меньший вклад [24], поэтому, как правило, не учитывается при расчете фотоэмиссии. Наконец, в-третьих, для возникновения поверхностного фотоэффекта напряженность электрического поля, возбуждающего электрон, должна иметь возможно большую составляющую, перпендикулярную поверхности металла. Очевидно, что это условие легче выполнить для наночастиц, чем для протяженных металлических пленок, на которые свет падает нормально или под небольшим углом. Таким образом, можно ожидать, что выход фотоэлектронов на единицу массы металла из наночастиц при возбуждении ЛПР окажется существенно больше, чем из протяженных металлических структур как, например, из слоев металлов, имеющихся в известных фотодетекторах [25, 26]. Повышение эффективности фотодетекторов за счет наночастиц поможет, в частности, повысить чувствительность фотодетекторов среднего и дальнего ИК диапазона, что является важной прикладной задачей [28,29]. Для количественной оценки фотоэмиссии из наночастиц и определения оптимальных параметров, при которых фотоэмиссия из наночастиц максимальна, необходимо рассчитать сечение фотоэмиссии из наночастиц, что является основной задачей настоящей работы. Расчет сечения поверхностной фотоэмиссии из металлических частиц с учетом возбуждения в них ЛПР, ее количественные оценки и, на основании этих оценок, заключение о том, что фотоэмиссия из наночастиц может быть существенно более эффективной, чем из сплошного слоя металла составляют один из основных результатов данной работы. Отметим, что создание микро- и наноструктур на поверхности металлических фотодетекторов для повышения эффективности фотоэмиссии известно и широко применяется, но эти структуры, в большинстве случаев, составляют единое целое с макроскопическими металлическими покрытиями, имеют с ними хороший электрический контакт, поэтому в таких структурах невозможно возбуждение ЛПР падающим на них полем и соответствующее увеличение фотоэмиссии. С другой стороны, использование плазмонных свойств наночастиц предлагалось для повышения эффективности фотодетекторов, однако фотоэмиссия из наночастиц при этом не рассматривалась [30]. Повышение эффективности фотоэмиссии из металлических наночастиц при возбуждении ЛПР наблюдалось экспериментально, и это предлагалось использовать для повышения эффективности фотодетекторов [31,32]. Вместе с тем, систематического теоретического исследования фотоэмиссии из наночастиц и определения сечения фотоэмиссии из наночастиц до настоящего времени не проводилось.

В настоящей работе будет выполнен расчет сечения фотоэмиссии из наночастиц, следуя классической теории, подробно изложенной в монографии [24]. Мы обобщаем результат этой теории для вероятности фотоэмиссии на случай, когда в явном виде учитываются поверхностные эффекты: изменения (скачки) электрического поля и массы электрона на поверхности частицы. Важный новый результат заключется в том, что учет упомянутых поврхностных эффектов в разы изменяет (в рассматриваемом случае - повышает) сечение фотоэмиссии. Учет поверхностных эффектов особенно важен для фотоэмиссии из наночастиц у которых относительно большая, в сравнении со сплошным слоем металла, площадь поверхности. Ниже обсуждается только фотоэмиссия электрона в вакуум или в однородную диэлектрическую среду, окружающую наночастицу, но рассматриваемый подход может быть применен и для решения и других, более сложных задач, связанных с переносом носителей в окрестности наночастиц. Например, о захвате свободного носителя частицей или фотоэмиссии из наночастицы, связанной с тунеллированием через потенциальный барьер [33], в том числе, когда металлическая частица находится вблизи поверхности полупроводника, но отделена от нее тонким (в т.ч. тунельным) диэлектрическим слоем. Последняя структура типична для задач о повышении эффективности солнечных элементов с помощью металлических наночастиц [16,17,18].

В разделе 2 находится общее выражение для амплитуды вероятности фотоэмиссии, при учете изменения электромагного поля и массы электрона при переходе электрона через границу сред. Для удобства, вывод необходимых для расчетов раздела 2 исходных формул [24] приводятся в приложениях. В разделе 3 дается явное выражение для амплитуды вероятности фотоэмиссии, когда граница металла представляется потенциалом в виде ступеньки. В разделе 4 определяется, с использованием результатов раздела 3, сечение фотоэмиссии электрона из наночастицы. В разделе 5 результаты раздела 4 используются для оценки сечения фотоэмиссии в кремний из сферических золотых наночастиц. Полученные результаты и направления дальнейших исследований фотоэмиссии из наночастиц обсуждаются в заключении.

## 2. Амплитуда вероятности фотоэмиссии при учете скчака электромагнитного поля на границе раздела сред.

В [24] по теории возмущений получено выражение для амплитуды вероятности $C_+(\infty)$ фотоэмиссии электрона, движущегося в среде (металле) вдоль оси $z$ перпендикулярной границе среды, под действием электромагнитного поля частоты $\omega$

$$C_+(\infty) = \frac{|e|m}{\hbar\omega W_1}\int_{-\infty}^{\infty} dz \left( E_m \frac{d\Psi_0}{dz}\Psi_{1-} + \frac{1}{2}\Psi_{1-}\Psi_0 \frac{dE_m}{dz}\right). \tag{1}$$

Здесь $e$ – (отрицательный) заряд, $m$ – масса электрона. Граница среды описывается одномерным потенциалом (барьером) $V(z)$, см. рис.1. Масса электрона, вообще говоря, меняется при выходе из металла, поэтому $m = m(z)$. В выражении (1), $\Psi_0$, $\Psi_{1\pm}$ – волновые функции электрона в состояниях с полной энергией $E_i$, $i = 0,1$ ниже и выше барьера, $E_m \equiv E/m(z)$, $E$ – амплитуда компоненты электрического поля, поляризованной вдоль оси $z$, $W_1$ определяется соотношениями, входящими в (55), (56) приложения 1. Окончательное выражение для вероятности фотоэмиссии в [24] получено в предположении, что электрическое поле $E$ постоянно вдоль оси $z$, но при пересечении границы сред это не так, т.к. на поверхности, из-за поверхностного заряда, происходит изменение нормальной компоненты поля[1], т.е. следует рассматривать $E(z)$.

Используя формулы, полученные в приложении 1 для $E = const$, выведем выражение для амплитуды вероятности фотоэмиссии для $E = E(z)$. Вывод удобно выполнить сначала для случая $m = const$, а затем обобщить его для случая $m(z)$.

---

[1]Граница металла на рис.1 – плавная функция $z$, поэтому и изменение электрического поля на границе – тоже плавная функция $z$.

## 2.1. Случай $m = const$, $E(z) \neq const$.

В этом случае в формуле (65) приложения 1 вместо $\hat{S}(d\Psi_0/dz)$ будет $\hat{S}[E(d\Psi_0/dz)]$. Учитывая, что при $m = const$ только слагаемое $\hat{T} = -[\hbar^2/(2m)](d^2/dz^2)$ не коммутирует с $E(z)$ в $\hat{S}$, можно записать (обозначая, здесь и далее, дифференцирование по $z$ соответствующим числом штрихов)

$$\hat{S}\left(E\frac{d\Psi_0}{dz}\right) = E\hat{S}\left(\frac{d\Psi_0}{dz}\right) - \frac{\hbar^2}{2m}(2E'\Psi_0^{\cdot} + E^{\cdot}\Psi_0). \tag{2}$$

Первое слагаемое в (2) даст вклад в $C_+(\infty)$ такой же, как и выражение (67) приложения 1, но $E(z)$ там должно находиться под интегралом. Во втором слагаемом в (2) запишем

$$2E'\Psi_0^{\cdot} + E^{\cdot}\Psi_0 = (E'\Psi_0)' + E'\Psi_0^{\cdot} = (E'\Psi_0)' + E'\frac{2m}{\hbar^2}[V(z) - E_0]\Psi_0, \tag{3}$$

где использовано то, что $\Psi_0$ удовлетворяет уравнению Шредингера (50) приложения 1 с энергией $E_0$. Подставляя (2) с учетом (3) в (63) приложения 1, добавляя к (63) приложения 1 второе слагаемое (1) и интегрируя слагаемое $\sim (E'\Psi_0')'$ по частям получаем общее выражение для амплитуды вероятности фотоэмиссии для случая $m = const$ и $E = E(z)$

$$C_+(\infty) = \frac{|e|}{W_1(\hbar\omega)^2}\int_{-\infty}^{\infty} dz\left\{-EV'\Psi_0\Psi_{1-} + E'\left[\frac{\hbar^2}{2m}\Psi_0'\Psi_{1-}' + \left(E_0 - V + \frac{\hbar\omega}{2}\right)\Psi_0\Psi_{1-}\right]\right\}. \tag{4}$$

Это выражение интегрируется, если $V' = V_0\delta(z)$ и $E' = (E_+ - E_-)\delta(z)$ где $E_{\pm}$ – значения $E$, соответственно, справа и слева от $z = 0$.

## 2.2. Случай $m \neq const$, $E(z) \neq const$.

Вернемся к выражению (1) с $m = m(z)$ и рассмотрим его первое слагаемое. Опуская, для краткости, постоянный множитель перед интегралом, запишем

$$\int_{-\infty}^{\infty} dz E_m\Psi_0'\Psi_{1-} = \frac{1}{\hbar\omega}\int_{-\infty}^{\infty} dz E_m\Psi_0'\hat{S}\Psi_{1-} = \frac{1}{\hbar\omega}\int_{-\infty}^{\infty} dz\Psi_{1-}\hat{S}(E_m\Psi_0'), \tag{5}$$

где $\hat{S} = H - E_0$, но в Гамильтониан $H$ входит оператор кинетической энергии $\hat{T}$, определяемый (50) приложения 1, с $m = m(z)$; в (5) использовано, что $\hat{S}$ – эрмитов оператор (см. приложение 3). Коммутируя $\hat{T}$ и $E_m$ находим

$$\hat{S}(E_m\Psi'_0) = E_m\hat{S}\Psi'_0 - \frac{\hbar^2}{2m(z)}(E'_m\Psi'_0)' + E'_m(E_0 - V)\Psi_0. \tag{6}$$

Можно получить, что $(\hat{S}\Psi_0)' = \hat{S}\Psi'_0 - (\hbar^2/2)[(1/m)'\Psi'_0]' + V'\Psi_0$ откуда, учитывая $\hat{S}\Psi_0 = 0$, следует

$$\hat{S}\Psi'_0 = (\hbar^2/2)[(1/m)'\Psi'_0]' - V'\Psi_0. \tag{7}$$

Подставляя (7) в (6) и затем в (5), интегрируя там по частям слагаемые $\sim (E'_m\Psi'_0)'$ и $\sim [(1/m)'\Psi'_0]'$ находим

$$\int_{-\infty}^{\infty} dz E_m\Psi'_0\Psi_{1-} = \frac{1}{\hbar\omega}\int_{-\infty}^{\infty} dz\left\{[E'_m(E_0 - V) - E_mV']\Psi_0\Psi_{1-} + \frac{\hbar^2 E'}{2m^2}\Psi'_0\Psi'_{1-}\right\}. \tag{8}$$

Подставляя (8) в (1) и заменяя $E_m = E/m$ приходим к общему выражению для амплитуды вероятности фотоэмиссии для $m = m(z)$ и $E = E(z)$

$$C_+(\infty) = \frac{|e|m}{W_1(\hbar\omega)^2}\int_{-\infty}^{\infty} \frac{dz}{m}(c_V + c_E + c_m), \tag{9}$$

где $c_V$, $c_E$ и $c_m$ описывают фотоэмиссию с учетом скачков потенциала, электромагнитного поля и эффективной массы электрона на границе раздела сред, соответственно,

$$c_V = -EV'\Psi_0\Psi_{1-}, \qquad c_E = E'\left[\frac{\hbar^2}{2m}\Psi'_0\Psi'_{1-} + \left(E_0 - V + \frac{\hbar\omega}{2}\right)\Psi_0\Psi_{1-}\right], \tag{10}$$

$$c_m = -\frac{Em'}{m}\left(E_0 - V + \frac{\hbar\omega}{2}\right)\Psi_0\Psi_{1-}.$$

В следующем разделе выражение (9) интегрируется для случая кусочно-линейных (ступенчатых) функций $V(z)$, $E(z)$ и $m(z)$, имеющих разрыв при $z = 0$ и постоянных при $z \neq 0$. Для них $V' = V\delta(z)$, $E' = (E_+ - E_-)\delta(z)$, $m' = (m_0 - m)\delta(z)$, где $E_\pm$, – значения $E$, соответственно, справа и слева от $z = 0$; $m$ и $m_0$ – эффективные массы электрона, соответственно, в металле и вне его. Величины $V(0) = V/2$, $E(0) = (E_+ + E_-)/2$, $m(0) = (m_0 + m)/2$.

# 3. Вероятность фотоэмиссии для границы в виде

## потенциала-ступеньки

### *3.1. Волновые функции в отсутствии возмущения.*

Найдем волновые функции электрона, который движется перпендикулярно границе металла, описываемой одномерным потенциальным барьером в виде ступеньки - рис.2. Электрон с зарядом $e$ и эффективной массой $m$ при $z < 0$ и $m = m_0$ при $z > 0$ движется в потенциале-ступеньке

$$V(z) = \begin{cases} 0 & z < 0 \\ V/2 & z = 0. \\ V & z > 0 \end{cases} \tag{11}$$

Гамильтониан электрона есть $H = H_-$ при $z < 0$ и $H = H_+$ при $z > 0$

$$H_- = -\frac{\hbar^2}{2m}\frac{d^2}{dz^2}, \qquad H_+ = -\frac{\hbar^2}{2m_0}\frac{d^2}{dz^2} + V. \tag{12}$$

Решая уравнение Шредингера $i\hbar(\partial\overline{\Psi}_{0+}/\partial t) = H_+\overline{\Psi}_{0+}$ находим волновую функцию $\overline{\Psi}_{0+} = \Psi_0 \exp[-i(E_0/\hbar)t]$ состояния электрона с полной энергией $E_0 < V$, падающего на барьер слева и отражающегосяот него

$$\Psi_0 = [\exp(ik_0 z) + A_0 \exp(-ik_0 z)]_{z<0} + [B_0 \exp(i\tilde{k}_0 z)]_{z>0}, \tag{13}$$

где волновые числа

$$k_0 = \frac{1}{\hbar}(2mE_0)^{1/2}, \qquad \tilde{k}_0 = \frac{1}{\hbar}[2m_0(E_0 - V)]^{1/2}. \tag{14}$$

Так как $V > E_0$, величина $(E_0 - V)^{1/2} = i(V - E_0)^{1/2}$ – чисто мнимая. Для удобства волновая функция $\Psi_0$ нормирована так, что коэфициент перед слагаемым $\exp(ik_0 z)$, описывающим исходное состояние электрона – падение на барьер из $z = -\infty$, выбран равным 1 (см. также замечание о нормировке $\Psi_{1-}$ после формулы (19)).

Коэффициенты $A_0$ и $B_0$ в (13) определяются из условий непрерывности при $z = 0$ (см. пояснение в приложении 3)

$$\Psi_0(z = -0) = \Psi_0(z = +0), \quad m^{-1}(\partial\Psi_0/\partial z)_{z=-0} = m_0^{-1}(\partial\Psi_0/\partial z)_{z=+0}, \tag{15}$$

которые эквивалентны уравнениям $1 + A_0 = B_0$ и $1 - A_0 = \theta_0 B_0$, откуда

$$A_0 = \frac{1-\theta_0}{1+\theta_0}, \quad B_0 = \frac{2}{1+\theta_0}, \quad \theta_0 = [(m/m_0)(1-V/E_0)]^{1/2}. \tag{16}$$

Аналогичным образом находим волновые функции $\bar{\Psi}_{1\pm} = \Psi_{1\pm} \exp\left[-i(E_1/\hbar)t\right]$ состояния электрона с полной энергией $E_1 > V$, где

$$\Psi_{1+} = [A_{1+}\exp(ik_1z) + B_{1+}\exp(-ik_1z)]_{z<0} + \exp(i\tilde{k}_1z)_{z>0}, \tag{17}$$

$$\Psi_{1-} = [A_{1-}\exp(i\tilde{k}_1z) + B_{1-}\exp(-i\tilde{k}_1z)]_{z>0} + \exp(-ik_1z)_{z<0}, \tag{18}$$

и действительные волновые числа

$$k_1 = \frac{1}{\hbar}(2mE_1)^{1/2}, \quad \tilde{k}_1 = \frac{1}{\hbar}[2m_0(E_1-V)]^{1/2}. \tag{19}$$

Согласно с ассимптотикой, принятой для решений уравнения Шредингера (50), $\Psi_{1+}$ соответствует расходящейся волне (т.е. электрон движется от барьера) для $z > 0$, а $\Psi_{1-}$ – расходящейся волне для $z < 0$. Для того, чтобы $C(\infty)$, определяемое (4), было бы амплитудой вероятности, нормировка $\Psi_{1+}$ должна быть выбрана таким образом, чтобы коэфициент перед слагаемым $\sim \exp(i\tilde{k}_1z)_{z>0}$, соответствующим расходящейся волне при $z > 0$, совпадал с коэфициентом перед слагаемым $\sim \exp(ik_0z)$, описывающим падение электрона на барьер из $z = -\infty$ в волновой функции (13), т.е. он должен быть равен 1, что и выбрано в (17). В противном случае $C(\infty)$ необходимо домножать на отношение коэфициентов при $\exp(i\tilde{k}_1z)_{z>0}$ и при $\exp(ik_0z)$. Нормировка волновой функции $\Psi_{1-}$ – произвольна, результат от нее не зависит, для симметрии она выбрана такой же, как и у волновой функции $\Psi_{1+}$.

Коэфициенты $A_{1\pm}$ и $B_{1\pm}$ определяются из условий непрерывности волновой функции при $z = 0$, аналогичных (15), которые приводят к уравнениям $A_{1\pm} + B_{1\pm} = 1$, $A_{1+} - B_{1+} = \theta_1$, и $\theta_1(B_{1-} - A_{1-}) = 1$, где $\theta_1 = [(m/m_0)(1-V/E_1)]^{1/2}$, откуда следует

$$A_{1+} = (1+\theta_1)/2, \quad B_{1+} = (1-\theta_1)/2, \quad A_{1-} = (\theta_1-1)/2\theta_1, \quad B_{1-} = (1+\theta_1)/2\theta_1. \tag{20}$$

Волновые функции (17) и (18) образуют фундаментальную систему решений в отсутствии возмущения, которая используется при нахождении решения неоднородного уравнения (53) приложения 1 с потенциалом-ступенькой (11).

### 3.2. Выражение для амплитуды вероятности

Вычисления с волновыми функциями (17) и (18) приводят к выражению

$$\left.\frac{W_1}{m}\right|_{z<0} = \left.\frac{W_1}{m_0}\right|_{z>0} = i\frac{k_1}{m}(1+\theta_1) \equiv i\left(\frac{k_1}{m} + \frac{\tilde{k}_1}{m_0}\right).$$ (21)

При $z = 0$, учитывая, что $\Psi'_0$, $\Psi'_{1-}$ — разрывные функции, так что $\Psi'_{0,1-}(0) = (1/2)[\Psi'_{0,1-}(-0) + \Psi'_{0,1-}(+0)]$, находим

$$\Psi_0\Psi_{1-} = \frac{2}{1+\theta_0}, \quad \Psi'_0\Psi'_{1-} = \frac{2\tilde{k}_0 k_1 \overline{m}^2}{m_0 m(1+\theta_0)},$$ (22)

здесь и далее $\overline{m} = (m_0 + m)/2$, $\Delta m = m_0 - m$, $\overline{E} = (E_+ + E_-)/2$, $\Delta E = E_+ - E_-$.

Подставляя (21) и (22) в (4) получаем, после преобразований, явное выражение для амплитуды вероятности фотоэмиссии

$$C_+(\infty) = \frac{2|e|m}{ik_1\overline{m}(\hbar\omega)^2(1+\theta_0)(1+\theta_1)}\left\{-V\overline{E} + \frac{\Delta E}{2}\left[E_1^{1/2} + (E_0 - V)^{1/2}\right]^2 - \frac{\overline{E}\Delta m}{2\overline{m}}(E_0 + E_1 - V)\right\}.$$ (23)

Для того, чтобы сравнить результат, следующий из (23), с известными и облегчить процедуру вычислений, удобно сначала рассмотреть случай $m = m_0$.

### 3.3. Вероятность фотоэмиссии для $m_0 = m$.

В этом случае, если выразить все в (23) через волновые числа, находим

$$C_+(\infty) = \frac{|e|k_0(ik_0 + |\tilde{k}_0|)}{2m\omega^2(k_1 + \tilde{k}_1)}(E_+ + E_-) - \frac{i|e|k_0}{(k_1 + \tilde{k}_1)\hbar\omega(k_0 + \tilde{k}_0)}\frac{(k_1 + \tilde{k}_0)^2}{k_1^2 - k_0^2}(E_+ - E_-).$$ (24)

Здесь первое слагаемое связано с поглощением фотона при столкновении электрона с барьером, а второе (при выводе которого использовано $\hbar\omega = \hbar^2(k_1^2 - k_0^2)/(2m)$) — с поглощением фотона за счет неоднородности электрического поля в области барьера. Первое слагаемое аналогично тому, которое получено в [24] в отсутствии скачка электрического поля[2] при $z = 0$. В пределе $V \to 0$, когда $\tilde{k}_{0,1} \to k_{0,1}$, второе слагаемое определяет вероятность перехода свободного электрона (т.е. при отсутствии барьера)

---

[2] Так как $E_\pm$ в (24) — амплитуды действительного электрического поля — см. определение (51), то результат следующий из (24) вдвое меньше, чем в [24]

$$C_+(\infty)_{V \to 0} \to -\frac{i\,|e|\,(E_+ - E_-)}{4\hbar\omega k_1} \frac{k_0 + k_1}{k_1 - k_0} \qquad (25)$$

между состояниями его непрерывного спектра под действием электрического поля со скачкобразно изменяющейся амплитудой при $z = 0$. Вероятность (25) можно получить независимо, например, из формул (62) приложения 1 или из (4).

Перепишем (24) в более компактном виде

$$C_+(\infty) = \frac{|e|\,k_0(ik_0 + |\tilde{k}_0|)}{2m\omega^2(k_1 + \tilde{k}_1)}\left[E_+ + E_- - \frac{\hbar^2(k_1 + i\,|\tilde{k}_0|)^2}{2mV}(E_+ - E_-)\right]. \qquad (26)$$

Из (26) следует, что при $E_+/E_- \sim 1$ слагаемое $\sim (E_+ - E_-)$, обусловленное скачком электрического поля, может быть порядка слагаемого $\sim (E_+ + E_-)$, обусловленного скачком потенциала. Действительно, второе слагаемое в квадратных скобках в (26) $\sim \hbar^2 k_1^2/(2m) = E_0 + \hbar\omega \geq V$. Учитывая граничные условия $\varepsilon_+ E_+ = \varepsilon_- E_-$, где $\varepsilon_\pm$ — диэлектрические функции справа и слева от барьера, можно записать

$$C_+(\infty) = \frac{|e|\,k_0(ik_0 + |\tilde{k}_0|)}{2m\omega^2(k_1 + \tilde{k}_1)}\left[\varepsilon_-/\varepsilon_+ + 1 - \frac{\hbar^2(k_1 + i\,|\tilde{k}_0|)^2}{2mV}(\varepsilon_-/\varepsilon_+ - 1)\right]E_-, \qquad (27)$$

где $E_-$ — компонента поля внутри частицы, нормальная к ее поверхности. Удобно представить $C_+(\infty)$, как функцию безразмерной переменной $x = (\hbar k_0)^2/(2mV) \equiv E_0/V$. Тогда

$$k_1 = \frac{\sqrt{2mV}}{\hbar}\left(x + \frac{\hbar\omega}{V}\right)^{1/2}, \quad \tilde{k}_1 = \frac{\sqrt{2mV}}{\hbar}\left(x + \frac{\hbar\omega}{V} - 1\right)^{1/2}, \quad |\tilde{k}_0| = \frac{\sqrt{2mV}}{\hbar}(1 - x)^{1/2}$$

и

$$|C_+(\infty)|^2 = C_0 U(x)\,|K_{dis}(x)|^2, \qquad (28)$$

здесь размерный коэффициент $C_0$ и безразмерная функция $U(x)$ есть

$$C_0 = \frac{2\,|e|^2\,|E_-|^2\,V}{m\hbar^2\omega^4}, \quad U(x) = \frac{x}{[(x + \hbar\omega/V)^{1/2} + (x + \hbar\omega/V - 1)^{1/2}]^2}, \qquad (29)$$

$K_{dis}(x)$ — фактор, описывающий фотоэмиссию за счет скачка электрического поля на границе раздела сред

$$K_{dis}(x) = \frac{1}{2}\left(\frac{\varepsilon_-}{\varepsilon_+} + 1\right) - \frac{1}{2}\left(\frac{\varepsilon_-}{\varepsilon_+} - 1\right)\left[\left(x + \frac{\hbar\omega}{V}\right)^{1/2} + i(1 - x)^{1/2}\right]^2. \qquad (30)$$

Если $\varepsilon_-/\varepsilon_+ = 1$, т.е. пренебрегается скачком электрического поля на границе раздела сред, то $K_{dis} = 1$. Для вещественного $\varepsilon_-/\varepsilon_+$ (что неверно для металлической частицы) $|K_{dis}(x)|^2$ приобретает простой вид

$$|K_{dis}(x)|^2 = \left[\frac{\hbar\omega}{2V}\left(\frac{\varepsilon_-}{\varepsilon_+}-1\right)-1\right]^2 + \left[\left(\frac{\varepsilon_-}{\varepsilon_+}\right)^2-1\right](1-x). \tag{31}$$

Для $\hbar\omega/V \to 0$ это выражение переходит в $|K_{dis}(x)|^2 = (\varepsilon_-/\varepsilon_+)(1-x)+x$.

### 3.4. Вероятность фотоэмиссии для $m \neq m_0$.

Выразим выражение (23) через переменную $x = (\hbar k_0)^2/(2mV) \equiv E_0/V$. Учитывая, что $k_1 = (\sqrt{2mV}/\hbar)(x+\hbar\omega/V)^{1/2}$, $\theta_0 = [(m/m_0)(1/x-1)]^{1/2}$, $\theta_1 = \{(m/m_0)[1-(x+\hbar\omega/V)^{-1}]\}^{1/2}$ находим, что в $|C_+(\infty)|^2$, определяемой (28)

$$U(x) = \frac{4r_m^2}{(r_m+1)^2} \cdot \frac{x}{[x+r_m(1-x)]\{(x+\hbar\omega/V)^{1/2}+[r_m(x+\hbar\omega/V-1)]^{1/2}\}^2}, \tag{32}$$

где        обозначено        отношение        масс        $r_m = m/m_0$,        и

$$K_{dis}(x) = \frac{1}{2}\left(1+\frac{\varepsilon_-}{\varepsilon_+}\right)\left[1+\frac{1-r_m}{1+r_m}\left(2x+\frac{\hbar\omega}{V}-1\right)\right] + \frac{1}{2}\left(1-\frac{\varepsilon_-}{\varepsilon_+}\right)\left[\left(x+\frac{\hbar\omega}{V}\right)^{1/2}+i(1-x)^{1/2}\right]^2. \tag{33}$$

Теперь фактор $K_{dis}$ описывает как влияние разрыва электромагнитного поля, так и разрыва значения эффективной массы электрона на границе металла при $z=0$, и выражение (32) для $U(x)$ тоже оказалось зависящим от скачка эффективной массы электрона. Чтобы вернуться к случаю, рассмотренному в [24], когда не принимались во внимание скачки массы электрона и амплитуды электромагнитного поля на границе металла, следует в (32), (33) положить $r_m = 1$ и $\varepsilon_- = \varepsilon_+$, что приведет к

$$K_{dis}(x)=1 \text{ и } U(x) = \frac{x}{\left[(x+\hbar\omega/V)^{1/2}+(x+\hbar\omega/V-1)^{1/2}\right]^2}.$$

Таким образом, вероятность фотоэмиссии электрона в случае границы в виде потенциала-ступеньки и при учете скачка эффективной массы электрона и амплитуды электромагнитного поля на границе раздела сред определяется (28), где коэфициент $C_0$ дается (29) а выражения для $U(x)$ и $K_{dis}(x)$ – определяются (32) и (33), соответственно. Ниже эти выражения будут использованы для расчета сечения фотоэмиссии из наночастицы.

# 4. Сечение фотоэмиссии для потенциала в виде ступеньки.

## 4.1. Выражение для сечения фотоэмиссии.

По определению, сечение $\sigma_{ph-em}$ фотоэмиссии из наночастицы

$$\sigma_{ph-em} = \frac{J_{ph-em}}{I},$$

(34)

где $J_{ph-em}$ – полный фототок из наночастицы в электронах в секунду, $I$ – интенсивность внешнего монохроматического поля, вызывающего фотоэмиссию, в месте, где находится частица, в фотонах через см$^2$ в секунду. Фототок из частицы

$$J_{ph-em} = \int_{surface} j ds = \int_{surface} j(\theta, \varphi, r) r dr \sin\theta d\theta d\varphi,$$

(35)

где $j$ – плотность фототока, в электронах через см$^2$ в сек., нормальная к поверхности частицы в точке поверхности, определяемой полярным углом $\theta$ азимутальным углом $\varphi$, и расстоянием $r$ от начала координат до поверхности частицы – рис.3.

Если, следуя [14; 21], перейти от одномерной модели движения электрона к трехмерной то, согласно формуле (2.30) из [24], плотность фототока $dj$ электронов, имевших энергию в интервале $E_0 \div E_0 + dE_0$ в области частицы (т.е. ниже барьера), поглотивших фотон и перешедших через барьер, будет

$$dj = \frac{\hbar \tilde{k}_{1z}}{m} |C_+|^2 \Theta[k_{0z}^2 + (2m/\hbar^2)(\hbar\omega - V)] dn_0,$$

(36)

где $\hbar \tilde{k}_{1z}/m$ – скорость таких электронов над барьером, $dn_0 \equiv 2 f_F(k_0) dk_{0x} dk_{0y} dk_{0z}/(2\pi)^3$ – число таких электронов,

$$f_F(k_0) = [1 + \exp\{[(\hbar k_0)^2/(2m) - \varepsilon_F]/k_B T\}]^{-1}$$

– функция расределения Ферми, $k_0$ – величина волнового вектора электрона до поглощения фотона, $k_0^2 = k_{0x}^2 + k_{0y}^2 + k_{0z}^2$, $k_{0x,y}$ – компоненты волнового вектора электрона, параллельные поверхности частицы, $\varepsilon_F$ – энергия Ферми металла, из которого состоит частица, $k_B$ – постоянная Больцмана, $T$ – температура. Так как

$2mE_0/\hbar^2 = k_0^2$, для барьера-ступеньки

$$\tilde{k}_{1z} = \sqrt{(2m/\hbar^2)(E_0 + \hbar\omega - V) - (k_{0x}^2 + k_{0y}^2)} \equiv \sqrt{k_{0z}^2 + (2m/\hbar^2)(\hbar\omega - V)} \tag{37}$$

– $z$-компонента волнового вектора электрона за барьером (выше, в одномерном случае обозначавшаяся, как $\tilde{k}_1$), $k_{0z}$ – компонента волнового вектора электрона внутри наночастицы, перпендикулярная ее поверхности, $|C_+|^2 \equiv |C_+(\infty)|^2$ – вероятность фотоэмиссии и $\Theta$ – тэта-функция. Плотность фототока $j$ электронов любой энергии

$$j = \frac{2}{(2\pi)^3} \int dk_{0x} dk_{0y} dk_{0z} f_F(\vec{k}_0) \frac{\hbar \tilde{k}_{1z}}{m} |C_+|^2 \Theta[k_{0z}^2 + (2m/\hbar^2)(\hbar\omega - V)]. \tag{38}$$

Учитывая, что в (38) только $f_F$ зависит от $k_{0x,y}$, там можно взять интеграл по $dk_{0x} dk_{0y}$, используя замену $k_{0x}^2 + k_{0y}^2 = \varrho^2$ и $\int_0^\infty dx(1 + e^x/b)^{-1} = \ln(1+b)$ ,

$$\int dk_{0x} dk_{0y} f_F(\vec{k}_0) = \pi \int_0^\infty \frac{d\varrho^2}{1 + \exp\{[\hbar^2(\varrho^2 + k_{0z}^2)/(2m) - \varepsilon_F]/k_B T\}} = \frac{2\pi m k_B T}{\hbar^2} \ln\{1 + e^{[\varepsilon_F - \hbar^2 k_{0z}^2/(2m)]/(k_B T)}\}. \tag{39}$$

Подставляя (36), (37), (39) в (38) и учитывая, что $|C_+|^2 \sim |E_-|^2$, согласно (28), (29) получаем плотность фотоэмиссионного тока в некоторой точке поверхности наночастицы

$$j = C_{emission} |E_-|^2, \tag{40}$$

$$C_{emission} = \frac{|e|^2 k_B T V^2}{\pi^2 \hbar^5 \omega^4} \int_{0,1-\hbar\omega/V}^{1} dx [1 + (\hbar\omega/V - 1)/x]^{1/2} \ln\left(1 + e^{\frac{\varepsilon_F - Vx}{k_B T}}\right) U(x) |K_{dis}(x)|^2,$$

где нижний предел интегрирования равен $0$, если $\hbar\omega > V$ и равен $1 - \hbar\omega/V$, если $\hbar\omega < V$; учтено, что $E_0 < V$ и следовательно $x < 1$; $U(x)$ и $K_{dis}(x)$ определяются, соответственно, (32) и (33).

Если пренебречь тепловым возбуждением электронов выше поверхности Ферми, т.е. перейти в (40) к пределу $T \to 0$ то, используя $e^{(\varepsilon_F - Vx)/(k_B T)} \to \infty$, вместо (40) можно записать

$$C_{emission} = \frac{|e|^2}{\pi^2} \frac{V^3}{\hbar^5 \omega^4} \int_{0,1-\hbar\omega/V}^{\varepsilon_F/V} dx [1 + (\hbar\omega/V - 1)/x]^{1/2} (\varepsilon_F/V - x) U(x) |K_{dis}(x)|^2, \tag{41}$$

при этом должно быть $\hbar\omega > V - \varepsilon_F$ и нижний предел интегрирования равен $0$ если $\hbar\omega > V$. Таким образом

$$J_{ph-em} = C_{emission} \int_{surface} |E_-|^2 \, ds, \tag{42}$$

здесь интеграл берется по поверхности частицы и $C_{emission}$ не зависит от координат точки поверхности; нормальная компонента поля $E_- = (\vec{E}_{int}\vec{n})$, где $\vec{E}_{int}$ – поле внутри частицы, $\vec{n}$ – вектор нормали к поверхности частицы в данной точке поверхности. Тангенциальные к поверхности частицы компоненты поля не влияют на вероятность фотоэмиссии. Движение электрона вдоль поверхности частицы под действием тангенциальных компонент поля влияет, в принципе, на функцию распределения электронов, но в относительно слабых полях, характерных для случаев фотоэмиссии, это влияние пренебрежимо мало, гораздо меньше, чем обычный нагрев частицы при поглощении электромагнитного поля. Поле внутри частицы связано с внешним, падающим на частицу полем $\vec{E}$ с соотношением $\vec{E}_{int} = \hat{F}(\vec{r})\vec{E}$, где $\hat{F}(\vec{r})$ – тензор. Для сфероидальных частиц, которые будут рассматриваться ниже, поле внутри частиц однородно и, следовательно, $\hat{F}$ – постоянный, не зависящий от $\vec{r}$. Для простоты предположим, что $\vec{E}$ параллелен одной из главных осей сфероидальной частицы, тогда $\vec{E}_{int} = F\vec{E}$, где $F$ – величина, не зависящая от $\vec{r}$. Для не сферических частиц $F$ зависит от того, какой из главных осей частиц параллелен $\vec{E}$. Таким образом,

$$J_{ph-em} = C_{emission} |F|^2 K_{geometry} |E|^2, \tag{43}$$

где $K_{geometry} = \int_{surface} (\vec{n}\vec{e})$, $\vec{e}$ – единичный вектор в направлении поляризации внешнего поля (см. рис.3). Учитывая формулу (34) и что интенсивность $I$ внешнего поля (в фотонах через см$^2$ в секунду)

$$I = \frac{1}{8\pi} \frac{cn_+ |E|^2}{hbar\omega},$$

находим сечение фотоэмиссии

$$\sigma_{ph-em} = \frac{8\pi\hbar\omega}{cn_+} C_{emission} |F|^2 K_{geometry}, \tag{44}$$

где $c$ – скорость света в вакууме, $n_+ = Re\sqrt{\varepsilon_+}$ – показатель преломления среды вне наночастицы.

*4.1. Параметры $F$ и $K_{geometry}$.*

Для сфероидальных частиц, согласно [37],

$$F = \frac{1}{1 + R_{dep} - iR_{rad}} \cdot \frac{\varepsilon_+}{\varepsilon_+ + (\varepsilon_- - \varepsilon_+)L} \qquad (45)$$

где второй сомножитель – результат вычислений в рамках квазистатического приближения [35], фактор $L = \frac{r^2}{2} \int_0^\infty \frac{du}{(u+r^2)^2(u+1)^{1/2}}$, аспектное отношение $r = a/c$, $a$ –

длина одной из двух одинаковых полуосей эллипсоидальной частицы, $c$ – длина

третьей полуоси – рис.3; первый сомножитель в (45), с помощью факторов $R_{dep}$ и

$R_{rad}$, соответственно, учитывает эффекты динамической деполяризации и

радиационных потерь [36]

$$R_{dep} = \frac{\varepsilon_- - \varepsilon_+}{\varepsilon_+ + (\varepsilon_- - \varepsilon_+)L}(A\varepsilon_+ y^2 + B\varepsilon_+^2 y^4), \quad R_{rad} = \frac{16\pi^3}{9}\frac{n_+^3}{r}\left(\frac{a}{\lambda}\right)^3 \frac{\varepsilon_- - \varepsilon_+}{\varepsilon_+ + (\varepsilon_- - \varepsilon_+)L} \qquad (46)$$

$$A = -0.4865L - 1.046L^2 + 0.848L^3, B = 0.01909L + 0.1999L^2 + 0.6077L^3,$$

где $y = \pi a/\lambda$, $\lambda$ – длина волны электромагнитного поля в вакууме. Фактор $R_{dep}$

характеризует отличие поля внутри наночастицы от однородного. Столкновения

электронов с поверхностью частицы приводят к тому, что диэлектрическая функция

$\varepsilon_-$ металла частицы оказывается отличной от диэлектрической функции $\varepsilon_{bulk}$

макроскопического куска того же металла. Это отличие можно учесть, следуя [3]

$$\varepsilon_- = \varepsilon_{bulk} + \frac{\omega_{pl}^2}{\omega^2 + i\omega\gamma_0} - \frac{\omega_{pl}^2}{\omega^2 + i\omega(\gamma_0 + iAv_F/a)}, \qquad (47)$$

где $\omega_{pl}$ и $\gamma_0$, соответственно, - плазменная частота и инкремент затухания из-за потерь, $v_F$ – скорость электронов у поверхности Ферми, $A$ – постоянная порядка 1, зависящая от формы частицы. Согласно [37],

$$K_{geometry} = \frac{\pi a^2}{r}\left[\frac{r}{1-r^2} + \frac{1-2r^2}{(1-r^2)^{3/2}}\arcsin(1-r^2)^{1/2}\right]. \qquad (48)$$

Таким образом, сечение фотоэмиссии определяется выражением (44), где коэфициент $C_{emission}$ определяется (40), факторы $F$ и $K_{geometry}$ – (45) и (48),

соответственно. Сечение фотоэмиссии можно сравнить с сечениями $\sigma_{abs}$ и $\sigma_{sc}$

поглощения и рассеяния света частицей [37]

$$\sigma_{abs} = \frac{8\pi^2 a^3 n_+}{3r\lambda} Im\alpha, \quad \sigma_{sc} = \frac{128\pi^5 a^6 n_+^4}{27r^2} |\alpha|^2, \quad \alpha = \frac{1}{1 + R_{dep} - iR_{rad}} \cdot \frac{\varepsilon_- - \varepsilon_+}{\varepsilon_+ + (\varepsilon_- - \varepsilon_+)L}. \quad (49)$$

В следующем разделе на основании этих результатов будут сделаны численные оценки сечений фотоэмиссии из металлических наночастиц.

## 5. Фотоэмиссия из золотых наночастиц в кремний

Вычислим, в качестве примера, сечение фотоэмиссии сферической золотой наночастицы, помещенной в кремний с р-типом легирования. Последний выбран в качестве оружения наночастицы потому, что работа выхода $\chi_e$ электрона из золота в кремий р-типа мала $\chi_e = 0.34$ эВ [38]. Так как энергия Ферми для золота $\varepsilon_F = 5.1$ эВ [38], высота барьера $V = \varepsilon_F + \chi_e = 5.44$ эВ. Масса электрона проводимости в золоте и в кремнии, соответственно, $m = 0.992 m_l \approx m_l$ и $m_0 = 0.25 m_l$ [39], где $m_l$ - масса свободного электрона в вакууме, таким образом, $r_m = 0.992 / 0.25 = 3.968$. Данные по диэлектрической функции золота $\varepsilon_{Au}$ взяты из [42]. Диэлектрическую функцию (47) металла наночастицы (золота) удобно записать, как функцию длины волны $\lambda$ ЭМП в вакууме:

$$\varepsilon_-(\lambda) = \varepsilon_{Au}(\lambda) + \left(\frac{\lambda}{\lambda_p}\right)^2 \left[\frac{1}{1 + i\lambda/\lambda_f} - \frac{1}{1 + (i\lambda/\lambda_f)(a_c/a + 1)}\right], \quad (49а)$$

при вычислениях взяты значения параметров: $\lambda_p = 0.142$ мкм, $\lambda_f = 55$ мкм, - они соответствуют лучшей аппроксимации

$$\varepsilon_{Au}(\lambda) \approx 12 + \left(\frac{\lambda}{\lambda_p}\right)^2 \frac{1}{1 + i\lambda/\lambda_f} \quad (49б)$$

в интересующей нас области $\lambda$ от 0.6 до 1.2 мкм, где лежит ЛПР рассматриваемой сферической золотой частицы в кремнии, $a_c = Av_f \lambda_f / (2\pi c_0)$, параметр, характеризующий столкновения электронов с поверхностью частицы, $A = 0.7$, $v_F = (2E_F |e| / m_0)^{1/2} \approx 1.3 \cdot 10^6$ м/с, и считается, что масса электрона проводимости в золоте практически совпадает с массой свободного электрона. На рис.4 для $a = 10$ нм показаны действительные и мнимые части $\varepsilon_{Au}(\lambda)$, ее аппроксимации по формуле (49б) – они достаточно близки, и $\varepsilon_-(\lambda)$, полученная по формуле (49а). Как видно, мнимая часть $\varepsilon_-(\lambda)$ существенно превосходит $\mathrm{Im}\left[\varepsilon_{Au}(\lambda)\right]$, что указывает на важность учета столкновений электронов с поверхностью наночастицы при оценке ее диэлектрической функции. Для диэлектрической функции $\varepsilon_+(\lambda)$ кремния, в который помещена золотая наночастица, воспользуемся аналитической аппроксимацией из [40]

$$\varepsilon_+(\lambda) = \varepsilon_\infty + \sum_{i=1}^{3} \frac{C_i}{1 - \left(\dfrac{1.242}{\lambda E_i}\right)^2 - i\dfrac{1.242}{\lambda E_i}\gamma_i} - F_1\,\chi_1^{-2}(\lambda)\ln\left[1 - \chi_1^2(\lambda)\right] - F_2\,\chi_2^{-2}(\lambda)\ln\frac{1-\chi_1^2(\lambda)}{1-\chi_2^2(\lambda)},$$

где $\chi_m(\lambda) = \left(\dfrac{1.242}{\lambda} + i\Gamma_m\right)\dfrac{1}{E_m}$ и $\varepsilon_\infty = 0.2$, $C_1 = 0.77$, $C_2 = 2.96$, $C_3 = 0.3$, $F_1 = 5.22$, $F_2 = 4$, $\gamma_1 = 0.05$, $\gamma_2 = 0.1$, $\gamma_3 = 0.1$, $E_1 = 3.38$, $E_2 = 4.27$, $E_3 = 5.3$, $\Gamma_1 = 0.08$, $\Gamma_2 = 0.1$.

На рис.5а показаны сечения поглощения и рассеяния (49) золотой наночастицы в кремнии в единицах $\pi a^2$. Как видно, в окрестности $\lambda = \lambda_{LPR} = 0.857$ мкм сечение поглощения более чем на порядок превосходит $\pi a^2$. На рис.5б изображен входящий в (44) фактор $|F|^2$ увеличения интенсивности поля в наночастице по сравнению с интенсивностью поля за ее пределами, F определяется (45). На частоте ЛПР интенсивность поля внутри частицы более чем в 150 раз превышает напряженность поля снаружи.

Сечение $\sigma_{ph-em}$ фотоэмиссии сферической золотой наночастицы радиуса $a = 10$ нм в кремнии, как функция длины волны падающего излучения, полученное по формуле (44), с использованием (40), (45) и (48), изображено на рис.6а вместе с сечениями поглощения $\sigma_{abs}$ и рассеяния $\sigma_{sc}$

Как видно из рис.6а, в максимуме ЛПР сечение фотоэмиссии достигает примерно половины величины геометрического сечения $\pi a^2$ и это составляет 4.2% максимальной величины $\sigma_{abs}$. Отношение $\sigma_{ph-em}/\sigma_{abs}$ в резонансе, т.е. при $\lambda = \lambda_{LPR}$, изображено на рис.6б и характеризует относительную долю энергии, поглощенной наночастицами, которая переходит в ток фотоэмиссии. Хотя эта доля не велика – несколько процентов, она оказывается заметно больше, чем для сплошного золотого слоя – см. ниже. Рис.6б показывает, что отношение $\sigma_{ph-em}/\sigma_{abs}$ почти линейно падает от 9% до 1% при увеличении радиуса наночастицы от 1 до 20 нм. Таким образом, фотоэмиссия более эффективна для маленьких частиц. Вместе с тем, само сечение поглощения мало при малых радиусах $a$ частицы (из-за уширения ЛПР, связанного со столкновениями электронов с поверхностью частицы), с увеличением радиуса наночастицы $\sigma_{abs}$ достигает максимума и дальше падает снова – из-за дефазировки и радиационных потерь – рис.6в. Таким образом, оптимальный радиус частицы можно оценить, скорее, исходя из величины тока фотоэмиссии – см. ниже.

При вычислении тока фотоэмиссии из ансамбля наночастиц существенную роль могут играть коллективные эффекты, в т.ч. взаимодействия частиц между собой через ЭМП. Детальное описание коллективных эффектов сложно и лежит за рамками настоящей работы, поэтому ниже ограничимся простыми и достаточно грубыми оценками влияния коллективных эффектов, оставив детальный анализ их влияния на фотоэмиссию на будущее. Исходя из того, что число поглощенных фотонов в слое наночастиц в единицу времени не может превышать числа поступающих извне на слой в единицу времени фотонов, можно получить, что $\sigma_{abs}/(\pi a^2)<1/\eta$ где $\eta$ - относительная поверхностная плотность наночастиц (доля площади поверхности, занятая наночастицами); т.е. $\sigma_{abs}$ должно уменьшается с ростом $\eta$. Это означает, что если для одной изолированной частицы $\sigma_{abs}>(\pi a^2)$, то с ростом $\eta$ происходит уширение ЛПР, что и наблюдается в экспериментах. По-видимому, уширение ЛПР из-за коллективных эффектов уменьшит также и фактор $F$ и сечение фотоэмиссии при узком спектре ЭМП, т.е. с $\lambda=\lambda_{LPR}$, но это не значит, что фотоэмиссия при широком спектре (напр. солнечном) обязательно уменьшится. Кроме того, в специальных случаях, в ансамблях наночастиц предсказаны и наблюдались узкие линии высокодобротных ЛПР [41], что означает, что эффективный фактор $F$ может оказаться весьма высок даже в ансамблях наночастиц, даже если $\sigma_{abs}$ не велико[3]. Таким образом, нельзя a-priori заключить, что коллективные эффекты заведомо уменьшат эффективность фотоэмиссии из наночастиц. Как отмечалось, детальное исследование влияния коллективных эффектов на фотоэмиссию из наночастиц достаточно сложно и выходит за рамки настоящей работы, поэтому сравним фотоэмиссию из наночастиц с фотоэмиссиией

---

[3] Известно, что большие локальные поля в ансамблях наночастиц могут возикать не только внутри, но и вне наночастиц, в областях между близкорасположенными частицами [1, Климов, стр.257]

из тонкого сплошного слоя золота в кремнии, пользуясь формулами, приведенными выше и полученными без учета коллективных эффектов.

Оценим фотоэмиссионный ток с единицы поверхности. Для тонкого сплошного слоя золота плотность $j_{ph-em}$ фотоэмиссионого оценим по формуле (40), где $E_-$ - компонента ЭМП, нормальная к поверхности металла. Допустим, для простоты, что имеет место скользящее падение ЭМП на поверхность металла, т.е. $E_- = E$ - амплитуде ЭМП.

Так как скользящее падение (т.е. угол падения ЭМП на поверхность металла $90^0$) реализовать трудно, мы получим заведомо завышенную оценку плотности фотоэмиссионого тока из сплошного слоя золота.

Допустим теперь, что в кремнии имеется слой сферических золотых наночастиц с относительной поверхностной плотностью $\eta$ и оценим поверхностную плотность тока фотоэмиссии из этого слоя под действием солнечного излучения. Спектр солнечного излучения, нормированный на 1, можно записать в виде $w_s(\lambda) = \dfrac{\lambda^{-4}/0.128}{\exp(2.616/\lambda)-1}$ где длина волны $\lambda$ - в микронах.

Поверхностная плотность тока $j_{ph-em}$ фотоэмиссии от слоя наночастиц, отнесенная к полной интенсивности $I$ солнечного излучения, будет

$$\frac{j_{ph-em}}{I} = \frac{\eta}{\pi a^2} \int w_s(\lambda)\sigma_{ph-em}(\lambda)d\lambda \ . \tag{49в}$$

На рис.7а показано отношение поверхностной плотности тока фотоэмиссии из слоя сферических золотых наночастиц к поверхностной плотности тока фотоэмиссии из сплошного слоя золота для $\eta = 0.3$ (т.е. 30% площади поверхности слоя покрыто наночастицами) – в зависимости от радиуса наночастицы, для спектральной области от 0.32 до 2 нм, где сосредоточено 80% энергии солнечного излучения.

На рис.7б – то же, для монохроматического ЭМП на длине волны локализованного плазмонного резонанса. Как видно из рис.7а, ток фотоэмиссии под действием солнечного излучения из слоя наночастиц в несколько раз превышает ток фотоэмиссии из сплошного слоя золота, при этом есть оптимальная величина радиуса наночастиц, при которой ток фотоэмиссии максимален. Без учета приповерхностной области частиц – что достаточно для поверхностного фотоэффекта.

коллективных эффектов на частоте плазмонного резонанса фотоэмиссионный ток и сечение фотоэмиссии из наночастиц превосходят соответствующие величины для сплошного золотого слоя в несколько сот раз – рис.7б. Для рассматриваемого случая, оптимальный радиус частиц – 10 нм. Разумеется, полученные результаты относятся к «внутренней» фотоэмиссии: вопрос об эффективности съема носителей во внешнюю цепь не рассматривается. Оценки, приведенные выше, как отмечалось, являются достаточно грубыми. Помимо коллективных эффектов они не учитывают, например, что наночастица и окружение (кремний) разделены потенциальным барьером, а не формируют потенциальную яму. В случае барьера возможно туннелирование носителей [33], кроме того, вероятность перехода через барьер при фотоэмиссии может быть выше, чем вероятость фотоэмиссии из потенциальной ямы. Учет потенциального барьера вместо потенциальной ямы может быть сделан по методике настоящей работы и приведет к увеличению сечения фотоэмиссии. Дополительный положительный, хотя небольшой, вклад в ток фотоэмиссии внесет объемная фотоэмиссия, которая не рассматривалась здесь, но которая может быть рассмотрена следуя, например, классической работе [34].

В настоящей работе, в отличие от [24], в конечном выражении для сечения фотоэмиссии приняты во внимание скачок нормальной компоненты электрического поля и скачок эффективной массы электрона проводимости на границе наночастицы и ее окружения. Оценим, на сколько принципиальны эти факторы при вычислении сечения фотоэмиссии из золотой наночастицы в кремний. Чтобы вернуться к результатам [24] следует сделать так, как сказано после (33). На рис.8 показаны сечения фотоэмиссии, если пренебрегать только скачком эффективной массы, т.е. $r_m = 1$, либо только скачком ЭМП, т.е. $\varepsilon_- / \varepsilon_+ = 1$, либо ими обоими т.е. $r_m = \varepsilon_- / \varepsilon_+ = 1$, а также сечение фотоэмиссии, полученное при $r_m \neq 1$ и $\varepsilon_- / \varepsilon_+ \neq 1$ - то же, что и кривая 3 на рис.6.

Согласно рис.8, учет как разрыва ЭМП так и разрыва эффективной массы в рассматриваемом примере существенно, в разы, изменяет величину сечения фотоэмиссии. Интересно, что максимальная величина сечения получается при учете всех разрывов, т.е. в данном случае скачки поля и массы помогают фотоэмиссии. Результаты, представленные на рис.8, указывают на то, что в общем случае, оба разрыва – и массы электрона и ЭМП существенны и их следует учитывать в теории.

Хотя в особых случаях, например, когда $\varepsilon_- / \varepsilon_+$ близко к 1 или фотоэмиссия происходит из золота в вакуум, так что к единице близко $r_m$, разрыв соответствующей величины несущественен. Интересно отметить, что скачок массы электрона, в отсутствии скачка поля, приводит к уменьшению сечения фотоэмиссии, отосительно результата [24], где скачки не учитывались (ср. кривые 2 и 3 на рис.8), скачок поля в отсутствии скачка массы увеличивает $\sigma_{ph-em}$ (ср. кривые 1 и 3 на рис.8), но одновременный учет всех скачков приводит к максимальному увеличению $\sigma_{ph-em}$ (ср. кривую 4 с остальными на рис.8). Такое «не аддитивное» влияние скачков ЭМП и массы электрона на сечение фотоэмиссии – следствие квантово-механической природы фотоэмиссии. Действительно, хотя слагаемые $c_V$, $c_E$ и $c_m$ описывающие, соответственно, фотоэмиссию с учетом скачков потенциала, электромагнитного поля и эффективной массы электрона на границе раздела сред входят в выражение (9) для амплитуды вероятности $C_+(\infty)$ фотоэмиссии аддитивным образом, в выражение (44) для $\sigma_{ph-em}$ входит $C_{emission} \sim |C_+(\infty)|^2$, и оказывается, что при вычислении $\sigma_{ph-em}$ может возникать существенная квантово-механическая интерференция вкладов от слагаемых $c_V$, $c_E$ и $c_m$, которая может приводить, как видно из примера, к возрастанию $\sigma_{ph-em}$ при учете обоих скачков, но к увеличению либо к уменьшению $\sigma_{ph-em}$, когда учитывается только скачок ЭМП либо только скачок эффективной массы электрона.

## 6. Заключение

В работе получена амплитуда вероятности и рассчитано сечение фотоэмиссии из металлических наночастиц. На примере сферических золотых наночастиц, находящихся в кремнии p-типа, показано, что на длине волны $\lambda = \lambda_{LPR}$, соответствующей возбуждению в наночастице локализованного плазмонного резонанса (ЛПР), сечение фотоэмиссии составляет несколько процентов от сечения поглощения наночастицы (которое в резонансе много большегеометрического сечения наночастицы); сечение фотоэмиссии составляет примерно половину геометрического сечения частицы – см. рис.6а. Таким образом, если относительная поверхностная плотность наночастиц в слое равна $\eta = 0.3$, то при $\lambda = \lambda_{LPR}$ в фотоэлектроны превратится примерно $0.5\eta$, т.е. 15% всех фотонов – это внутренний КПД фотоэмиссии на частоте ЛПР. Фотоэмиссионный ток от слоя наночастиц при $\lambda = \lambda_{LPR}$ может быть на 2 порядка больше, чем от сплошного слоя золота. При облучении широким солнечным спектром наночастицы дают в несколько раз больший фотоэмиссионный ток, чем сплошной слой металла – рис.7. Существует оптимальный радиус металлической наночастицы, при котором достигается максимум фотоэмиссионного тока, см. рис.7. Увеличение фотоэмиссии из наночастиц в сравнении со случаем сплошного слоя металла, обусловлено, как возрастанием интенсивности поля в наночастице при возбуждении ЛПР (рис.5) так и существенной частью поверхности наночастицы, нормальной к поляризации падающего на частицу поля.

Обобщена теория [24] фотоэмиссии из металлов: при расчете сечения фотоэмиссии учтены разрывы эффективной массы электронов проводимости и нормальной компоненты напряженности электрического поля на границе металла и внешней среды. Получены соответствующие аналитические выражения и, на примере фотоэмиссии из золотых наночастиц в кремний, показано, что учет данных разрывов существенно (в несколько раз) изменяет величину сечения.

При расчетах не принимался во внимание объемный фотоэффект, который увеличит фототок, а потенциальная кривая на границе раздела наночастица-внешняя среда аппроксмировалась прямоугольной потенциальной ямой. Учет потенциального барьера, через который возможно тунеллирование, приведет к увеличению фотоэмиссионного тока. Расчет с учетом более сложных потенциальных кривых может быть выполнен прямым обобщением методики данной работы. В дальнейшем, при переходе от отдельных наночастиц к их ансамблям, при исследовании фотоэмиссии следует учитывать коллективные эффекты при взаимодействии наночастиц через ЭМП. Последние отразятся только на величине фактора F, описывающего поле внутри частицы. Используя подход настоящей работы, может быть проанализирован захват носителей металлическими наночастицами. Результаты работы могут быть использованы для создания новых высокочуствительных фотодетекторов и фотопреобразователей светового излучения в электрическую энергию.

# Приложение 1. Амплитуда вероятности фотоэмиссии.

## П1.1. Общее решение по теории возмущений.

Предполагается, что длина волны Де-Бройля, соответствующая электрону проводимости металла, много меньше неоднородности поверхности металлической наночастицы, связанной с кривизной ее границы, т.е. для электрона граница металла наночастицы является поверхностью, перпендикулярой оси $z$. Среда (металл), окружающая электрон, поглощающий ЭМП, рассматривается классически и задается одномерным потенциалом $V(z)$, $V(\infty) = V_0$, $V(-\infty) = 0$, изображенным на рис.1. Такое приближение при анализе фотоэффекта из металлов является общепринятым, хотя металл и его граница состоят из отдельных атомов и, строго говоря, должны рассматриваться, как квантовые объекты. Электрон в металле – квазичастица, поэтому при переходе через границу эффективная масса $m$ электрона изменяется, т.е. $m = m(z)$.

Следуя [24], вычислим вероятность перехода электрона через границу при поглощении фотона электромагнитного поля, когда последнее рассматривается, как возмущение, т.е. предполагается, что вероятность фотоэмиссии окажется мала по сравнению с единицей. Рассмотрим сначала случай, когда электрон движется перпендикулярно границе вдоль оси $z$, учет движения электрона вдоль границы выполняется независимо от анализа движения электрона вдоль оси $z$ (см. раздел 4). Волновая функция $\Psi(z) e^{-(iE_i/\hbar)t}$ электрона в состоянии с полной энергией $E_i$ определяется из уравнения Шредингера

$$E_i \Psi = H \Psi, \quad H = T + V(z), \quad T = -\frac{\hbar^2}{2} \frac{d}{dz}\left(\frac{1}{m}\frac{d}{dz}\right), \tag{50}$$

где $T$ – оператор кинетической энергии для частицы с переменной массой $m(z)$ (см. пояснение в Приложении 2). Решения уравнения (50) считаются известными. Имеется два линейно независимых решения $\Psi_{i+}$ и $\Psi_{i-}$, образующих фундаментальную систему решений, т.е. общее решение (50) есть линейная комбинация $\Psi_{i+}$ и $\Psi_{i-}$. Ассимптотика $\Psi_{i\pm}$ – плоские волны, движущиеся от границы: при $z \to -\infty$ $\Psi_{i-} \to C_{i-} \exp(-ik_i z)$ а при $z \to \infty$ - $\Psi_{i+} \to C_{i-} \exp(i\tilde{k}_i z)$, где $C_{i\pm}$ – нормировочные постоянные, $k_i$ и $\tilde{k}_i$ – волновые числа электрона при $z < 0$ и $z > 0$, соответственно.

Допустим, на электрон действует переодическое во времени возмущение – электромагнитное (световое) поле

$$U(z, d/dz)\cos(\omega t) = (1/2)U(z, d/dz)(e^{-i\omega t} + e^{i\omega t}). \qquad (51)$$

Будем рассматривать только процессы поглощения электроном фотона энергии $\hbar\omega$, процессами с испусканием фотона можно пренебречь, так как при нормальных условиях число электронов с кинетической энергией превышающей $\hbar\omega$ ничтожно мало. В этом случае в выражении (51) следует сохранить только слагаемое $\sim e^{-i\omega t}$, соответствующее получению электроном энергии от возмущающего поля.

Рассмотрим процесс, когда под действием возмущения $U$ электрон в состоянии $\Psi_0$, движущийся положительном направлении оси $z$ со стороны $z < 0$, переходит из состояния с энергией $E_0 < V_0$ в состояние с энергией $E_1 = E_0 + \hbar\omega > V_0$. Волновая функция такого электрона будет

$$\Psi = \Psi_0 e^{-i(E_0/\hbar)t} + \psi e^{-i(E_1/\hbar)t}. \qquad (52)$$

Подставляя волновую функцию (52) в уравнение Шредингера

$$i\hbar(\partial\Psi/\partial t) = [H + U(z, d/dz)e^{-i\omega t}]\Psi$$

и используя теорию возмущений, т.е. пренебрегая слагаемым $\sim U(z, d/dz)\psi$, получаем уравнение для $\psi$

$$E_1\psi = H\psi + f(z), \quad f(z) = (1/2)U(z, d/dz)\Psi_0. \qquad (53)$$

Уравнение (53) – линейное неоднородное уравнение второго порядка. Его решение $\psi$, как известно [43] стр.272, выражается через решения $\Psi_{1+}$ и $\Psi_{1-}$ уравнения $E_1\psi = H\psi$, как

$$\psi = C_+(z)\Psi_{1+} + C_-(z)\Psi_{1-}, \qquad (54)$$

где

$$C_\pm(z) = \mp\frac{2}{\hbar^2}\int_{\mp\infty}^{z} dz' \frac{m(z')\Psi_{1\mp}(z')f(z')}{W(\Psi_{1-}\Psi_{1+})} + A_\pm, \quad W(\Psi_{1-}\Psi_{1+}) = \Psi_{1-}\frac{d\Psi_{1+}}{dz} - \Psi_{1+}\frac{d\Psi_{1-}}{dz}, \qquad (55)$$

и $A_+$, $A_-$ – постоянные, определяемые из граничных условий. Согласно ассимптотике волновых функций $\Psi_{1\pm}$ граничные условия – $C_+ \to 0$ при $z \to -\infty$, $C_- \to 0$ при $z \to \infty$. Эти условия выполняются при $A_\pm = 0$. Учитывая, что $\Psi_{1\pm}$ удовлетворяют уравнению Шредингера (50) с одним и тем же $E_i = E_1$, получаем

$$\frac{d}{dz}\left(\frac{W}{m}\right) = \Psi_{1-}\frac{d}{dz}\left(\frac{1}{m}\frac{d\Psi_{1+}}{dz}\right) - \Psi_{1+}\frac{d}{dz}\left(\frac{1}{m}\frac{d\Psi_{1-}}{dz}\right) = -\frac{2}{\hbar^2}\left(\Psi_{1-}\hat{T}\Psi_{1+} - \Psi_{1+}\hat{T}\Psi_{1-}\right) = 0,$$

т.е. $W(\Psi_{1-}\Psi_{1+})/m$ не зависит от $z$ и следовательно

$$C_\pm(z) = \mp\frac{m}{\hbar^2 W_1}\int_{\mp\infty}^{z} dz' \Psi_{1\mp}(z')U(z',d/dz')\Psi_0(z'), W_1/m \equiv W(\Psi_{1-}\Psi_{1+})/m = const, \qquad (56)$$

где $f(z)$ записано в виде (53).

Соотношения (52), (54) и (56) определяют, в первом порядке теории возмущений, волновую функцию электрона, проходящего над потенциальным барьером под действием переодического возмущения. Величина $|C_+(\infty)|^2$ — вероятность перехода электрона через потенциальный барьер, которую надо определить.

## *П1.2. Явный вид возмущения и амплитуды вероятности (56).*

Если электрон переходит через потенциальный барьер под действием электромагнитного поля, тогда в уравнении Шредингера следует заменить оператор импульса $\hat{p} = -i\hbar(d/dz)$ на $\hat{p} - (e/c)A$ и добавить туда $e\varphi$, так что пренебрегая слагаемыми $\sim A^2$

$$U\Psi_0 = \frac{i\hbar e}{2c}\left[\frac{d}{dz}\left(\frac{A}{m}\Psi_0\right) + \frac{A}{m}\frac{d\Psi_0}{dz}\right] + e\varphi\Psi_0, \qquad (57)$$

где $e$ — заряд электрона (отрицательный), $c$ — скорость света в вакууме, $A$ — векторный потенциал электромагнитного поля в среде, $\varphi$ — скалярный потенциал электромагнитного поля. Предположим пока, что векторный потенциал имеет только одну $z$-компоненту: $A_z = A$, т.е. электрическое поле поляризовано вдоль оси $z$ (см. также замечание после формулы (42)). Найдем явный вид интеграла $\int_{-\infty}^{\infty} d\tilde{z}\Psi_{1-}(\tilde{z})U(\tilde{z},d/d\tilde{z})\Psi_0(\tilde{z})$, который определяет искомую вероятность перехода электрона через барьер. Вычислим сначала вклад от первого слагаемого в (57) (обозначая, для краткости, дифференцирование по $z$ штрихом)

$$\int_{-\infty}^{\infty} dz\, \Psi_{1-} \left[ \frac{d}{dz}\left( \frac{A}{m}\Psi_0 \right) + \frac{A}{m}\frac{d\Psi_0}{dz} \right] = \int_{-\infty}^{\infty} dz\, \frac{A}{m}\left[ \Psi_{1-}\frac{d\Psi_0}{dz} - \Psi_0\frac{d\Psi_{1-}}{dz} \right] \equiv$$

$$\int_{-\infty}^{\infty} d\left[ \int_{-\infty}^{z} A(\tilde{z})d\tilde{z} \right]\left[ \frac{\Psi_{1-}}{m}\frac{d\Psi_0}{dz} - \frac{\Psi_0}{m}\frac{d\Psi_{1-}}{dz} \right] = \int_{-\infty}^{\infty} dz\left[ \int_{-\infty}^{z} A(\tilde{z})d\tilde{z} \right]\left[ \Psi_0\left( \frac{\Psi_{1-}}{m} \right)' - \Psi_{1-}\left( \frac{\Psi_0}{m} \right)' \right] =$$

$$\frac{2}{\hbar^2}\int_{-\infty}^{\infty} dz\left[ \int_{-\infty}^{z} A(\tilde{z})d\tilde{z} \right]\left( \Psi_1\hat{T}\Psi_0 - \Psi_0\hat{T}\Psi_{1-} \right) = -\frac{2\omega}{\hbar}\int_{-\infty}^{\infty} dz\,\Psi_0\Psi_{1-}\left[ \int_{-\infty}^{z} A(\tilde{z})d\tilde{z} \right]. \qquad (58)$$

Здесь сначала проинтегрировано по частям слагаемое $\sim d(A\Psi_0)/dz$ и предполагается, что $(A\Psi_0\Psi_{1-})_{z=\pm\infty} = 0$; затем получившееся выражение проинтегрировано по частям и опять предполагается, что слагаемые, не входящие в интеграл, обращаются в $0$ при $z = \pm\infty$; использовано уравнение Шредингера (50) и $E_1 - E_0 = \hbar\omega$. Заменяя во втором слагаемом в (57) $\varphi \equiv \int_{-\infty}^{z} d\tilde{z}(d\varphi/d\tilde{z})$ и подставляя затем (57) в (56), используя при этом с результат (58), находим

$$C_+(\infty) = -\frac{em}{\hbar^2 W_1}\int_{-\infty}^{\infty} dz\,\Psi_0\Psi_{1-}\int_{-\infty}^{z}\left( \frac{-i\omega A}{c} + \frac{d\varphi}{d\tilde{z}} \right)d\tilde{z}. \qquad (59)$$

Выражение, стоящее под интегралом в круглых скобках в (3) есть $-E$, где $E$ – $z$-компонента амплитуды электрического поля, т.е.

$$C_+(\infty) = -\frac{|e|m}{\hbar^2 W_1}\int_{-\infty}^{\infty} dz\,\Psi_0\Psi_{1-}\int_{-\infty}^{z} E d\tilde{z}, \qquad (60)$$

где явно выделен знак заряда электрона[4]. Из уравнения Шредингера (50) следует

$$\frac{d}{dz}\left( \frac{W_{10}}{m} \right) = \frac{2\omega}{\hbar}\Psi_0\Psi_{1-}, \quad W_{10} = \Psi_{1-}\frac{d\Psi_0}{dz} - \Psi_0\frac{d\Psi_{1-}}{dz}. \qquad (61)$$

Используя (61), интегрируя (60) по частям и полагая, как обычно, что слагаемые, не входящие в интеграл, обращаются в 0 при $|z| \to \infty$, получаем

$$C_+(\infty) = \frac{|e|m}{2\hbar\omega W_1}\int_{-\infty}^{\infty} E(z)\frac{W_{10}(z)}{m(z)}dz, \qquad (62)$$

– амплитуду вероятности перехода электрона через барьер под действием монохроматического электромагнитного поля. Преобразуем (62), следуя [24], стр.42.

---

[4]Если принять $\hbar^2 W_1/(2m) = -1$, как это сделано в [24], то формула (60) отличается множителем 1/2 от (2.41) из [24]. Множитель 1/2 возник из-за соответствующего

Обозначая $E_m \equiv E/m$, подставляя $W_{01}$ из (61) в (62), интегрируя после этого по частям второе слагаемое в (62), считая при этом, как и выше, что то, что не входит под интеграл обращается в 0 при $z = \pm\infty$, получаем выражение (1). Удобно преобразовать (1) так, чтобы слагаемые под интегралом содержали производные $V$, $E$ или $m$ – тех величин, которые испытывают скачок на границе раздела сред. Ниже такое преобразование выполнено для случая $E = const$ и $m = const$ и получен результат [24], а в разделе 2 оно выполнено для общего случая $E \neq const$ и $m \neq const$.

*П1.3. Амплитуда вероятности фотоэмиссии при $E = const$ и $m = const$.*

В этом случае

$$C_+(\infty) = \frac{|e|E}{\hbar\omega W_1} \int_{-\infty}^{\infty} dz \frac{d\Psi_0}{dz} \Psi_{1-}. \tag{63}$$

Обозначим $\hat{S} = H - E_0$, так что $\hat{S}\Psi_0 = 0$ и $\hat{S}\Psi_{1-} = \hbar\omega\Psi_{1-}$. Подставляя $\Psi_{1-} = (\hbar\omega)^{-1}\hat{S}\Psi_{1-}$ в (63) получаем

$$C_+(\infty) = \frac{|e|E}{W_1(\hbar\omega)^2} \int_{-\infty}^{\infty} dz \frac{d\Psi_0}{dz} (\hat{S}\Psi_{1-}). \tag{64}$$

Так как $\hat{S}$ (так же как и оператор Гамильтона $H$) является эрмитовым оператором (см. Приложение 3), то (см. [43] стр.422)

$$\int_{-\infty}^{\infty} dz \frac{d\Psi_0}{dz} (\hat{S}\Psi_{1-}) = \int_{-\infty}^{\infty} dz \Psi_{1-} \left( \hat{S} \frac{d\Psi_0}{dz} \right). \tag{65}$$

Учитывая, что по определению $\hat{S}\Psi_0 = 0$ и, следовательно,

$$0 = \frac{d}{dz}\left( \hat{S}\Psi_0 \right) \equiv \hat{S}\frac{d\Psi_0}{dz} + \frac{dV(z)}{dz}\Psi_0 \tag{66}$$

и $\hat{S}(d\Psi_0/dz) = -(dV(z)/dz)\Psi_0$. Таким образом, вместо (64) получаем

$$C_+(\infty) = -\frac{|e|E}{W_1(\hbar\omega)^2} \int_{-\infty}^{\infty} dz \frac{dV}{dz} \Psi_0\Psi_{1-}. \tag{67}$$

Если $dV/dz = V_0\delta(z)$, то

---

множителя 1/2 в(51), которого нет в [24].

$$C_+(\infty) = -\frac{|e|EV_0}{W_1(\hbar\omega)^2}\Psi_0(0)\Psi_{1-}(0), \tag{68}$$

что совпадает с соответствующей формулой из [24].

## Приложение 2

Получим выражения для некоторых операторов для случая массы электрона, зависящей от координаты. Рассматривается одномерное движение электрона.

1. Оператор кинетической энергии. Оператор импульса есть $p = -(i/\hbar)d/dz$, соответственно, оператор скорости - $v = p/m = -(i/\hbar m)d/dz$. Следовательно, оператор кинетической энергии

$$T \equiv \frac{mv^2}{2} = -\frac{\hbar^2}{2}\frac{d}{dz}\left(\frac{1}{m}\frac{d}{dz}\right).$$

2. Для электрона в электромагнитном поле с единственной отличной от $0$ $z$-компонентой векторного потенциала $A$, оператор квазиимпульса $p_q = p - (e/c)A$. Соответственно, оператор "квазискорости" $v_q = (1/m)[p - (e/c)A]$. Следовательно, оператор "квазикинетической энергии" (пренебрегая слагаемыми $\sim A^2$)

$$T_q \equiv \frac{mv_q^2}{2} = T - \frac{ie\hbar}{2c}\left[\frac{d}{dz}\left(\frac{A}{m}\right) + \frac{A}{m}\frac{d}{dz}\right],$$

где слагаемое $\sim \hbar$ – возмущение.

## Приложение 3

Оператором в $\hat{S}$ является $\tilde{T} \equiv d/dz(qd/dz)$, $q(z) \equiv m^{-1}(z)$. Покажем, что оператор $\tilde{T}$ эрмитов на области $-\infty < z < \infty$ для функций $f(z)$ которые, вместе с их первыми производными, обращаются в 0 при $|z| \to \infty$. Производную по $z$ обозначаем, дкак обычно, штрихом

$$\int_{-\infty}^{\infty}dzf_1\tilde{T}f_2dz \equiv \int_{-\infty}^{\infty}dzf_1(qf_2')'dz = \int_{-\infty}^{\infty}f_1d(qf_2) = f_1qf_2'\Big|_{-\infty}^{\infty} - \int_{-\infty}^{\infty}qf_2f_1'dz =$$

$$-\int_{-\infty}^{\infty}qf_1df_2 = -f_2qf_1'\Big|_{-\infty}^{\infty} + \int_{-\infty}^{\infty}dzf_2(qf_1')'dz = \int_{-\infty}^{\infty}dzf_2\tilde{T}f_1dz.$$

Предполагается, что функции выше удовлетворяют требуемым условиям.

# Приложение 4

Интегрируя уравнение Шредингера (50) один раз находим

$$\frac{1}{m(z)}\psi'(z) = \frac{2}{\hbar^2}\int_{-\infty}^{z}[V(\tilde{z})-E]\Psi(\tilde{z})d\tilde{z}.$$

Отсюда

$$\frac{1}{m(+0)}\psi'(+0) = \frac{1}{m(-0)}\psi'(-0) + \frac{2}{\hbar^2}\lim_{\delta\to 0}\int_{-\delta}^{\delta}[V(\tilde{z})-E]\Psi(\tilde{z})d\tilde{z}.$$

Так как выражение под интегралом не содержит $\delta$-функций, сам интеграл является непрерывной функцией, следовательно предел есть $0$, откуда следует второе из граничных условий (15).

Проценко Игорь Егеньевич 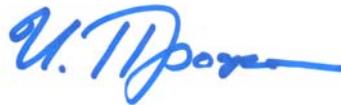

Усков Александр Васильевич

# Литература

**Подписи к рисункам**

Рис.1 Потенциальный барьер, в котором движется электрон.

Рис.2 Потенциальный барьер в виде ступеньки. Указан разрыв электрического поля при $z = 0$ – в месте разрыва потенциала и диэлектрические функции $\varepsilon_-$ металла и $\varepsilon_+$ окружения частицы.

Рис.3 Плотность фототока $j$ в точке поверхности, определяемой углами $\theta$ и $\varphi$, сфероидальной наночастицы, возбуждаемой внешним электромагнитным полем амплитуды $E$. Длины двух полуосей частицы одинаковы и равны $a$, длина третьей полуоси – $c$.

Рис.4. Мнимая 1 и действительная 2 части диэлектрической функции золота согласно [42] - пунктир и согласно (49a) – сплошная линия. Горизонтальная линия из точек обозначает 0.

Рис.5 (а) Сечения поглощения $\sigma_{abs}$ (сплошная линия) и рассеяния $\sigma_{sc}$ (пунктир) сферической золотой наночастицы в кремнии. (б) Фактор увеличения интенсивности поля в наночастице.

Рис.6 (а) Сечение фотоэмиссии золотой наночастицы (кривая 3) и ее сечения поглощения (кривая 1) и рассеяния (кривая 2) в единцах $\pi a^2$ - геометрического сечения наночастицы. (б) отношение $\sigma_{ph-em} / \sigma_{abs}$ в максимуме ЛПР для различных радиусов сферических золотых наночастиц; (в) сечение поглощения сферической золотой наночастицы в зависимости от ее радиуса.

Рис.7 (а) Отношение поверхностной плотности тока фотоэмиссии из слоя сферических золотых наночастиц, с поверхностной плотностью 30%, к поверхностной плотности тока фотоэмиссии из тонкого сплошного слоя золота в кремнии – в зависимости от радиуса $a$ частиц. (б) То же – на длине волны $\lambda_{LPR}$

Рис.8. Сечения фотоэмиссии. Пренебрегается скачком: эффективной массы (кривая 1); ЭМП (кривая 2); и эффективной массы и ЭМП (кривая 3) – результат, согласно [24]; при учете всех скачков (кривая 4).

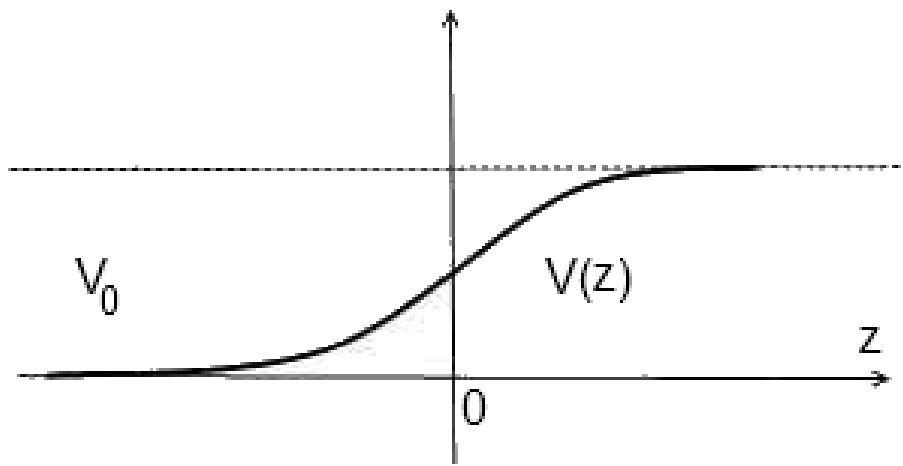

Рис 1

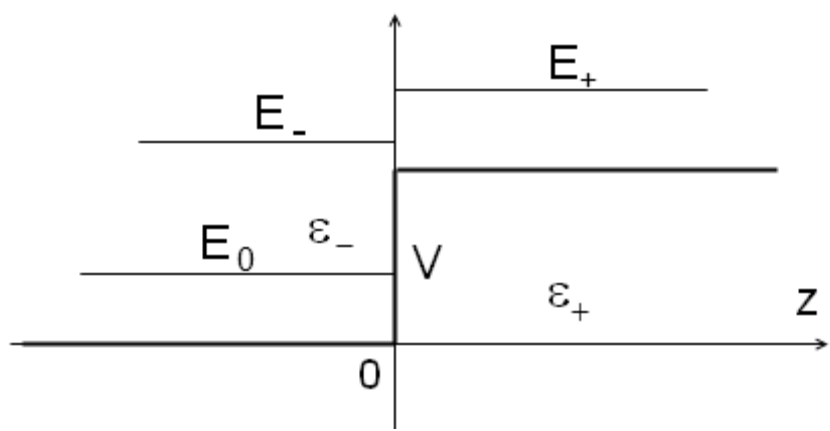

Рис.2

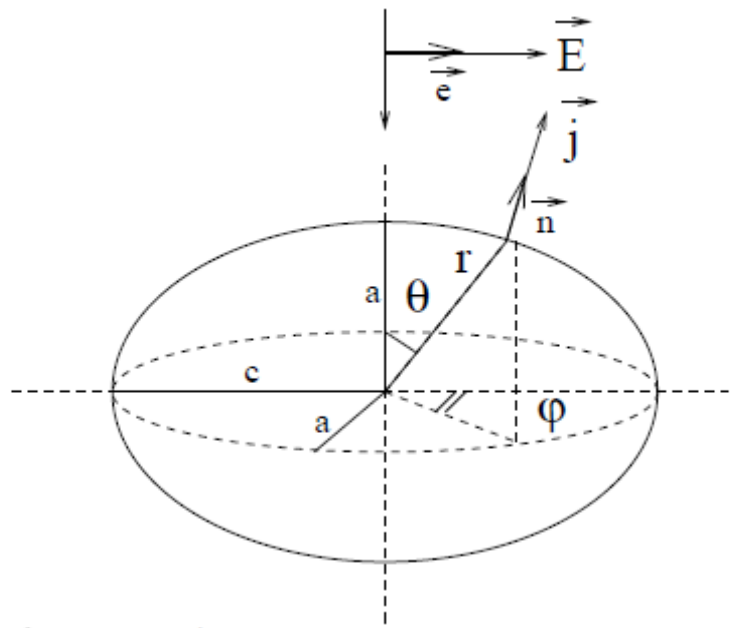

Рис.3

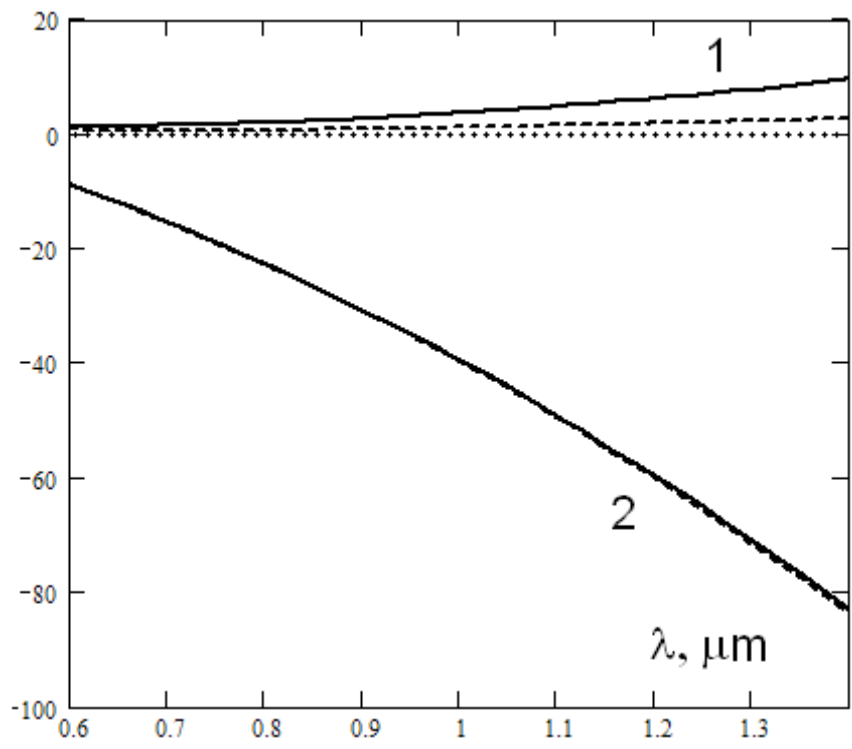

Рис.4

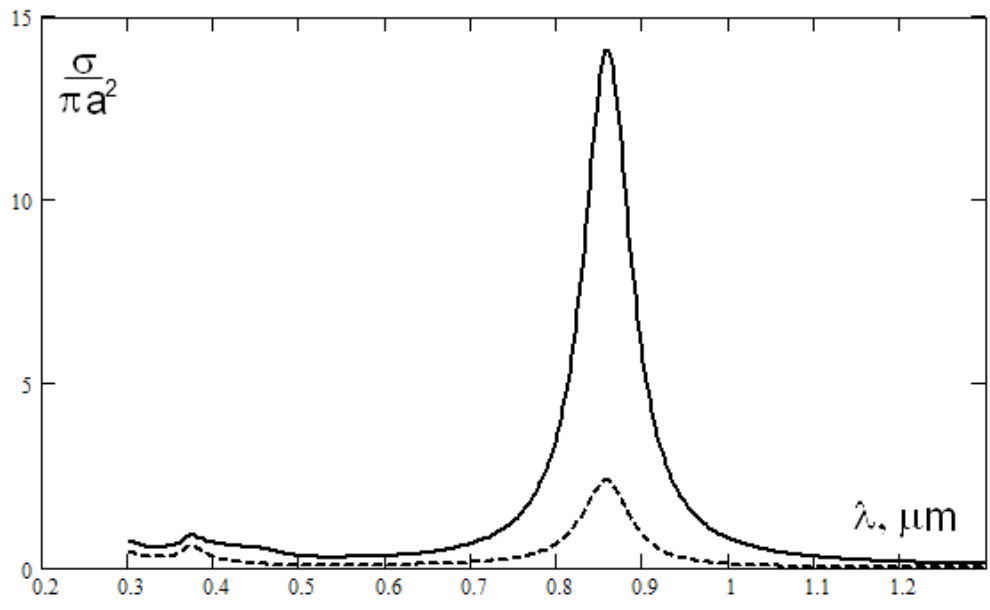

Рис.5а

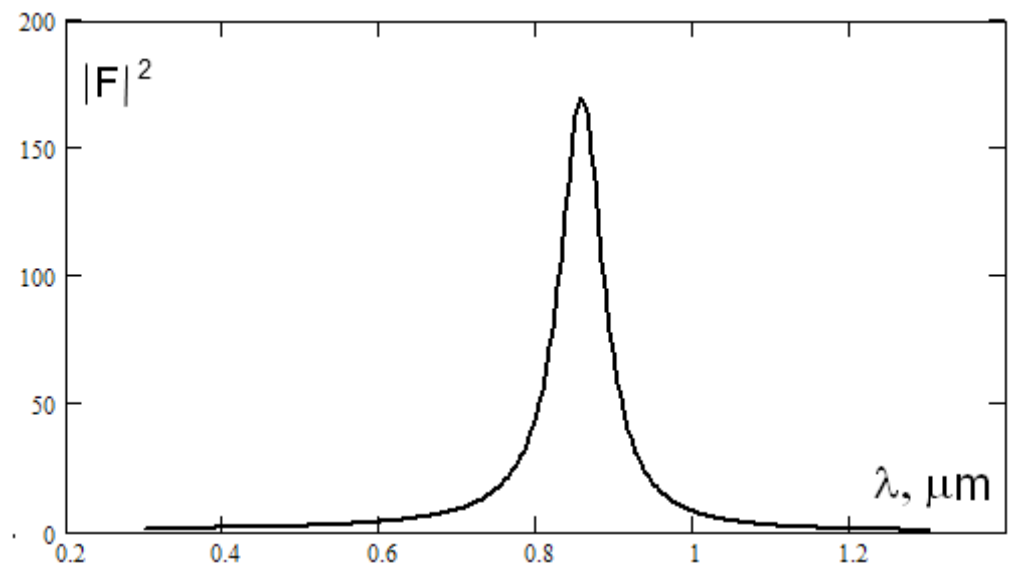

Рис.5б

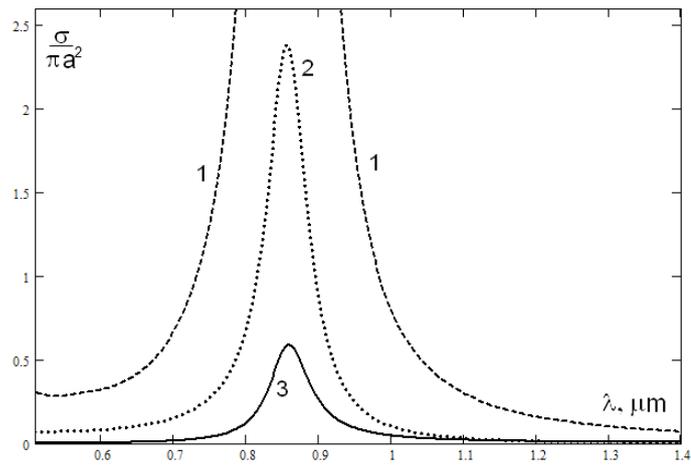

Рис.6а

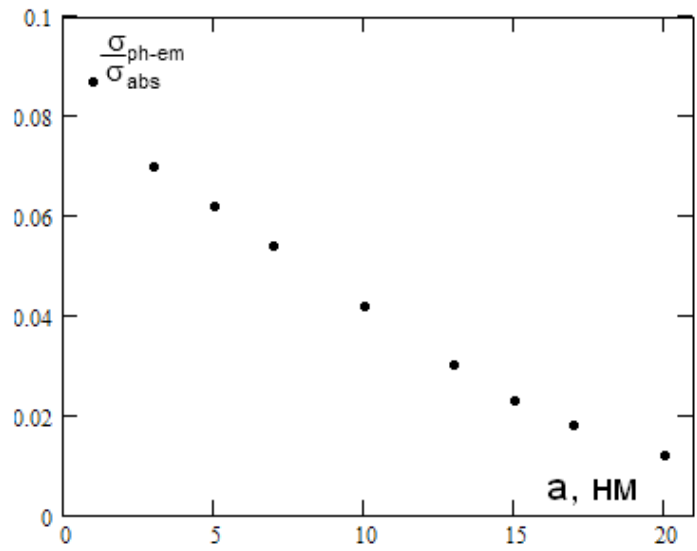

Рис. 6б

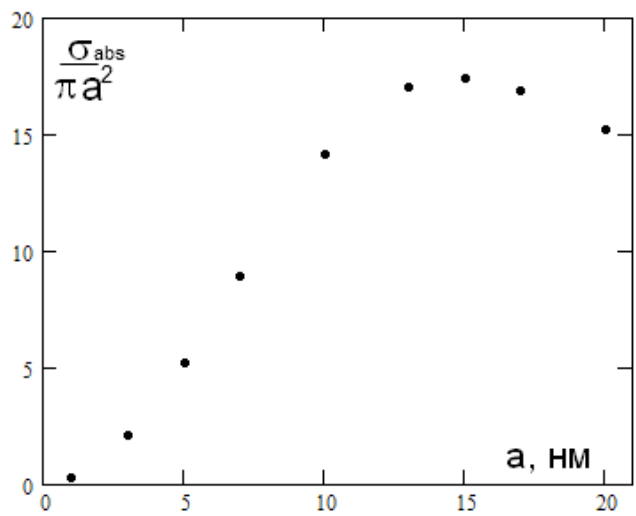

Рис.6в

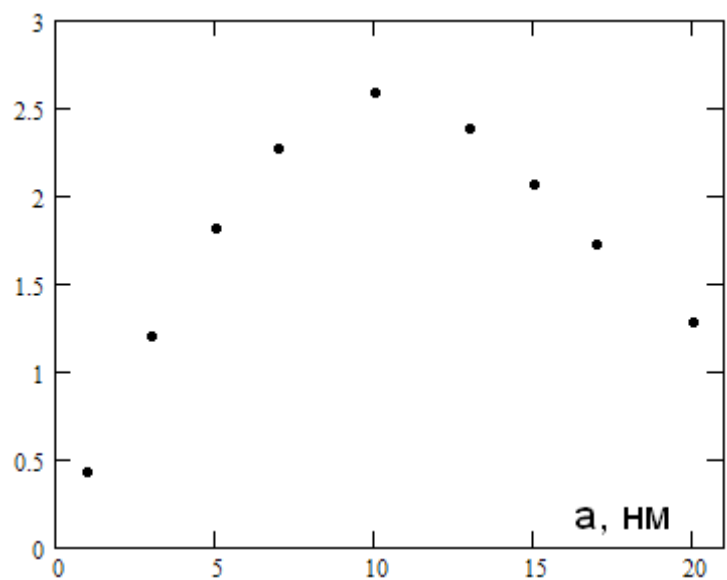

Рис.7а

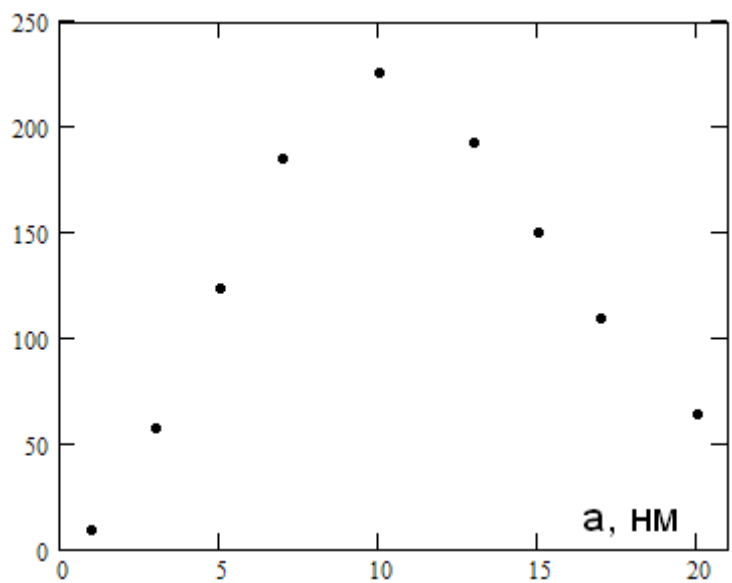

Рис.7б

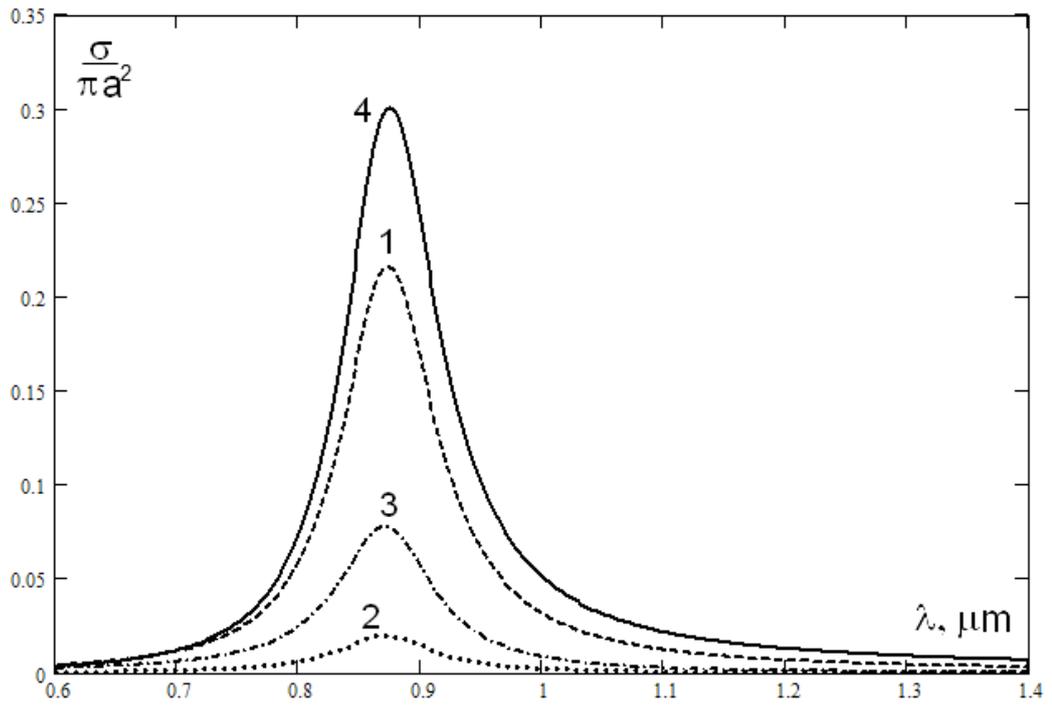

Рис.8

PHOTOEMISSION FROM METAL NANOPARTICLES

I.E.Protsenko, A.V.Uskov

*Lebedev Physical Institute of RAS, 119991, Moscow, Leninskiy prospect 53*

*Plasmonics LTD, 109382, Moscow, Nijnie Polia street, 52/1.*

protsenk@sci.lebedev.ru; alexusk@sci.lebedev.ru

Key words: localized plasmon resonance, nano-patricles, surface photoemission

# Abstract

A.Brodsky and Yu.Gurevitch approach is discussed and generalized for photoemission from metal nano-particles taking into account the excitation of localized plasmon resonance (LPR) and changes of electromagnetic field (EMF) and conduction electron mass in the metal – environment interface. New result is the increase of photo-emission current several time respectively to the case of continues metal film due to increase of intensity of EMF near the surface of nanoparticles and also due to surface phenomena mentioned above. Results can be applied for development new photodetectors, photo energy converters (solar cells) and for more studies of photoemission from metal nanoparticles.

**Авторы:**

Проценко Игорь Евгеньевич, (для переписки) Москва, 127055, ул.Лесная д.45 кв.30; тел: 8-903-502-30-13, 499-973-36-11; protsenk@gmail.com; к.ф.-м.н., с.н.с. Физического института им П.Н.Лебедева РАН, 119991, Москва, Ленинский Проспект, 53; зав. лабораторией ТФКЭ ООО «Плазмоника», 109382, Москва, ул. Нижние поля 52/1.

Усков Александр Васильевич, Москва, 117594, проезд Одоевского 7, к.6, кв.748; тел: +7-91549373937, +7-495-4236396; alexusk@lebedev.ru; в.н.с. Физического института им П.Н.Лебедева РАН, 119991, Москва, Ленинский Проспект, 53; ООО «Плазмоника», 109382, Москва, ул. Нижние поля 52/1.